\documentclass[10pt,twocolumn,letterpaper]{article}

\usepackage{iccv}
\usepackage{times}
\usepackage{epsfig}
\usepackage{microtype}
\usepackage{graphicx}
\usepackage{subfigure}
\usepackage{booktabs} 
\usepackage{amsmath,amssymb,amsfonts}
\usepackage{algorithm}
\usepackage{textcomp}
\usepackage{xcolor}
\usepackage{soul}
\usepackage{makecell}
\usepackage{ctable}
\usepackage{booktabs} 
\usepackage{multirow}
\usepackage{thm-restate}
\usepackage{sistyle}
\SIthousandsep{,}
\usepackage{calligra}
\usepackage{enumitem}
\usepackage{subfigure}
\usepackage{footmisc} 
\usepackage{cite}
\usepackage{url}

\usepackage[breaklinks=true,bookmarks=false]{hyperref}

\iccvfinalcopy 

\def\iccvPaperID{****} 

\ificcvfinal\fi

\begin{document}

\title{Color Image Denoising Using The Green Channel Prior}

\author{Zhaoming Kong\\
South China University of Technology\\
Guangzhou, China\\
{\tt\small kong.zm@mail.scut.edu.cn}
\and
Xiaowei Yang\\
South China University of Technology\\
Guangzhou, China\\
{\tt\small xwyang@scut.edu.cn}
}

\maketitle
\ificcvfinal\thispagestyle{empty}\fi

\begin{abstract}

Noise removal in the standard RGB (sRGB) space remains a challenging task, in that the noise statistics of real-world images can be different in R, G and B channels. In fact, the green channel usually has twice the sampling rate in raw data and a higher signal-to-noise ratio than red/blue ones. However, the green channel prior (GCP) is often understated or ignored in color image denoising since many existing approaches mainly focus on modeling the relationship among image patches. In this paper, we propose a simple and effective one step GCP-based image denoising (GCP-ID) method, which aims to exploit the GCP for denoising in the sRGB space by integrating it into the classic nonlocal transform domain denoising framework. Briefly, we first take advantage of the green channel to guide the search of similar patches, which improves the patch search quality and encourages sparsity in the transform domain. Then we reformulate RGB patches into RGGB arrays to explicitly characterize the density of green samples. The block circulant representation is utilized to capture the cross-channel correlation and the channel redundancy. Experiments on both synthetic and real-world datasets demonstrate the competitive performance of the proposed GCP-ID method for the color image and video denoising tasks. The code is available at github.com/ZhaomingKong/GCP-ID.

\end{abstract}
\vspace{-12.8pt}
\section{Introduction}
The rapid development of digital cameras has largely facilitated the recording of colorful and dynamic objects, preserving and presenting rich information of real scenes. Inevitably, image data are contaminated by noise to varying degrees during acquisition and transmission. Therefore image denoising plays an indispensable role in modern imaging systems. Meanwhile the growth in the number and size of images also poses a greater demand on noise removal in terms of both effectiveness and efficiency. \\
\indent The target of noise removal is to recover a clean image $\mathcal{X}$ from its noisy observation $\mathcal{Y} = \mathcal{X} + \mathcal{N}$, where $\mathcal{N}$ is usually modeled as additive white Gaussian noise (AWGN) with standard deviation (std) $\sigma$. Image denoising enjoys a long history and early works focus on handling single-channel images \cite{buades2005review, aharon2006k, zoran2011learning}. When the input is an RGB image, naively filtering each channel separately will lead to unsatisfactory results since the spectral correlation among RGB channels is ignored \cite{xu2017multi}. To improve the channel-by-channel algorithm, two solutions are mainly adopted. The first strategy is to apply a decorrelation transform along the RGB dimension, and then handle each channel of the transformed space independently. For example, the representative CBM3D method \cite{dabov2007color} considers the luminance-chrominance (\textit{e.g.}, YCbCr) space as a less correlated color space. An alternative strategy is to jointly denoise RGB channels simultaneously by concatenating RGB patches, and applying traditional transforms such as singular value decomposition (SVD) \cite{xu2017multi, jha2010denoising} and wavelet filters \cite{tian2023multi, ruikar2011wavelet} or deep neural networks (DNNs) such as convolution neural networks (CNNs) \cite{zhang2017beyond, zhang2018ffdnet} and transformers \cite{zamir2022restormer, liang2021swinir}.\\
\indent Despite the outstanding performance of existing methods, they are restricted to directly treating RGB patch as a whole or assigning different weights to each channel. In addition, the main focus of many related works lies in the modeling of relationship among image patches by imposing various regularization terms and solving complex optimization problems \cite{chang2017hyper, mahdaoui2022image}, which may drastically increase computational burden and fail to exploit the difference of noise statistics in RGB channels. It is noticed that to reconstruct a full-resolution color image from sensor readings, a digital camera generally goes through the image demosaicing process \cite{menon2011color} based on the Bayer color filter array (CFA) pattern \cite{chung2008lossless}, which measures the green channel at a higher sampling rate than red/blue ones. Such prior knowledge is known as the green channel prior (GCP), however the GCP is commonly used in processing and applications of raw sensor data \cite{liu2020joint, guo2021joint, zhang2022joint}, which are not always available. Therefore, it is interesting to investigate if the GCP can be leveraged to guide denoising in the sRGB space. \\
\indent Motivated by the fact that the green channel has a higher signal-to-noise ratio (SNR) than red/blue ones in most natural images, in this paper, we present a GCP-based image denoising (GCP-ID) method for color images. The proposed GCP-ID follows the traditional nonlocal transform domain framework due to its simplicity, effectiveness and adaptiveness \cite{kong2023comparison, elad2023image}. The performance of traditional patch-based denoisers is often affected by two decisive factors: nonlocal similar patch search and proper representation for local image patches. To improve the classic paradigm, GCP-ID utilizes the green channel to guide the search of similar patches because the green channel is less noisier and preserves better image structure and details. Then the similar patches are stacked into a group and the group-level relationship can be captured by performing principal component analysis (PCA) along the grouping dimension \cite{dabov2009bm3d, zhang2010two}. In order to model the density of green samples and the importance of the green channel, each RGB image patch is reformulated into RGGB array based on the Bayer pattern. The spectral correlation among RGB channels is further exploited with block circulant representation. Eventually, the patch-level redundancy can be characterized by the tensor-SVD (t-SVD) transform \cite{kilmer2011factorization, kilmer2013third}. Our contributions can ne summarized as follows:
\vspace{-5pt}
\begin{itemize}
  \item To explicitly take advantage of the GCP for sRGB data, we present a novel denoiser GCP-ID by integrating it into the classic patch-based denoising framework. \vspace{-3.8pt}
  \item We propose to reformulate image patches as RGGB arrays and utilize block circulant representation to capture their spectral correlation. The formulation is simplified as a one step t-SVD transform.
      \vspace{-3.8pt}
  \item Experiments demonstrate the competitive performance of GCP-ID for both real-world color image and video denoising tasks. Besides, we study how the GCP may be extended to other imaging techniques such as Multispectral/Hyperspectral imaging (MSI/HSI).
\end{itemize}
\vspace{-8.8pt}
\section{Notations and Preliminaries}
\indent In this paper, vectors and matrices are denoted by boldface lowercase letters $\mathbf{a}$ and capital letters $\mathbf{A}$, respectively. Tensors \cite{kolda2009tensor} are denoted by calligraphic letters, e.g., $\mathcal{A}$. Given an $N$th-order tensor $\mathcal{A} \in \mathbb{R}^{I_1\times I_2\times\cdots\times I_N}$, the Frobenius norm of a tensor $\mathcal{A} \in \mathbb{R}^{I_1\times I_2\times\cdots\times I_N}$ is defined as $\|\mathcal{A}\|_F = \sqrt{\sum_{i_1=1}...\sum_{i_N=1}\mathcal{A}_{i_1...i_N}^2}$. The $n$-mode product of a tensor $\mathcal{A}$ by a matrix $\mathbf{U}\in \mathrm{R}^{P_n\times I_n}$ is denoted by $\mathcal{A}\times _n\mathbf{U} \in \mathbb{R}^{I_1 \cdots I_{n-1} P_n I_{n+1} \cdots I_N}$. \\
\indent The product between matrices can be generalized to the product of two tensors according to t-product \cite{kilmer2013third}. Specifically, the t-product `$*$' between two third-order tensors $\mathcal{A} \in \mathbb{R}^{N_1 \times N_2 \times N_3}$ and $\mathcal{B} \in \mathbb{R}^{N_1 \times N_4 \times N_3}$ is also a third order tensor $\mathcal{C} \in \mathbb{R}^{N_1 \times N_4 \times N_3}$ with $\mathcal{C} = \mathcal{A} * \mathcal{B}$ computed by
\begin{equation}\label{Equ_t_product}
  bcirc(\mathcal{C}) = bcirc(\mathcal{A}) bcirc(\mathcal{B})
\end{equation}
where $bcirc$ is the block circulant operation that reshapes a third order tensor into a block circulant matrix \cite{kilmer2013third}. Directly handling block circulant matrices is time consuming. Eq. (\ref{Equ_t_product}) can be computed efficiently in the Fourier domain
\begin{equation}\label{Equ_t_SVD_Fourier_domain}
  bdiag(\mathcal{C}_{FFT}) = bdiag(\mathcal{A}_{F}) bdiag(\mathcal{B}_{F})
\end{equation}
where $bdiag$ is the block diagonal representation, $\mathcal{A}_{F}$ is obtained by performing the fast Fourier transform (FFT) along the third mode of $\mathcal{A}$ via $\mathcal{A}_F = \mathcal{A} \times _3\mathbf{W}_{FFT}$, and $\mathbf{W}_{FFT}$ is the fast Fourier transform (FFT) matrix. \\
\indent The tensor-SVD or t-SVD of a third-order tensor can then be defined as the t-product of three third-order tensors
\begin{equation}\label{Equ_t-SVD}
  \mathcal{A} = \mathcal{U} * \mathcal{S} * \mathcal{V}^T
\end{equation}
\vspace{-20pt}
\section{Related Works}
\begin{figure}[htbp]
\graphicspath{{Figs/}}
  \centering
  \includegraphics[width=2.699in, height = 2.099in]{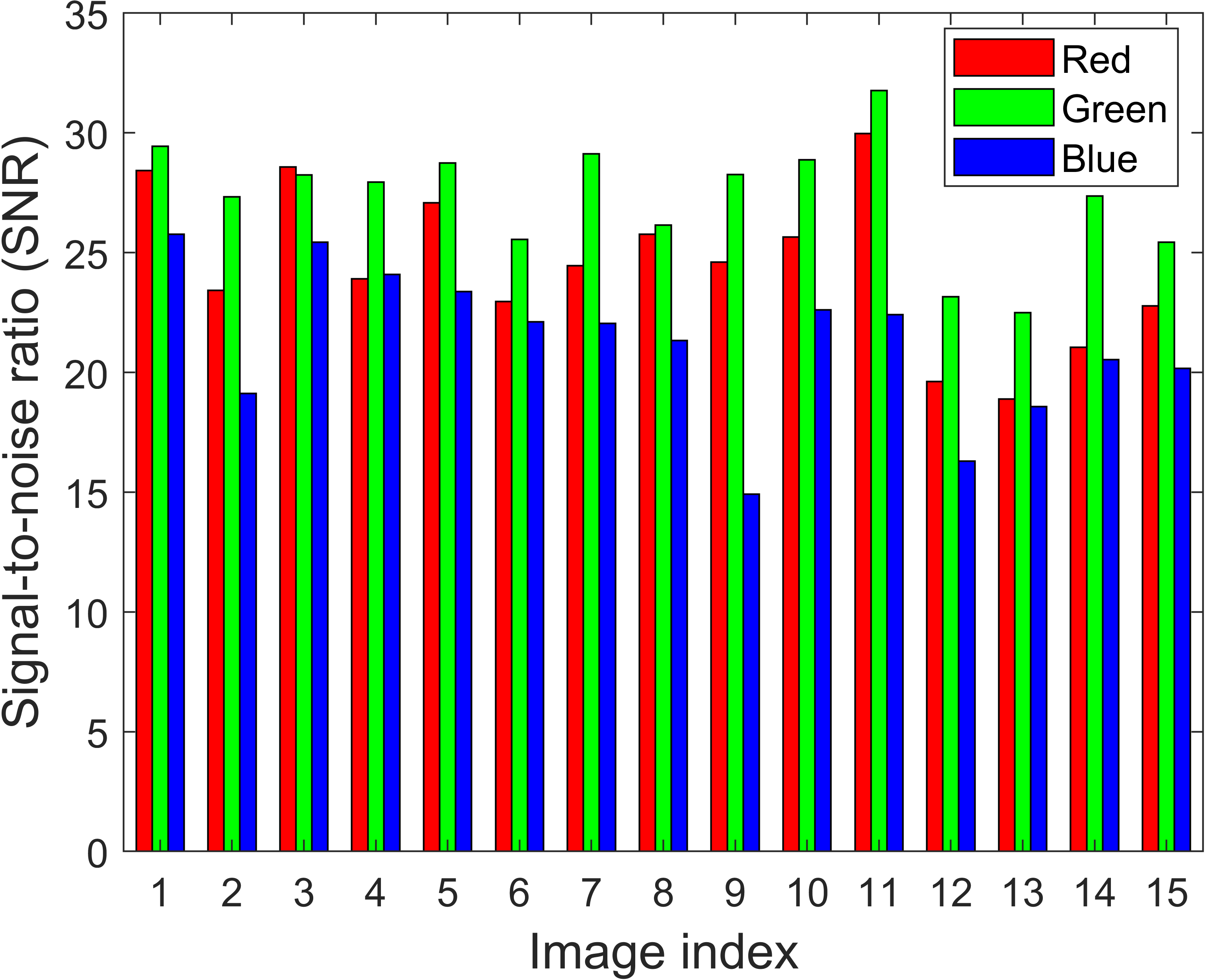}\\
  \caption{Signal-to-noise ratio (SNR) in RGB channels of 15 real-world images from the CC15 dataset.}
  \label{Fig_SNR_comparison}
\end{figure}

\indent \textbf{Nonlocal transform domain framework.} Natural images are known to have the nonlocal self-similarity (NLSS) characteristics, which refers to the fact that an image patch often has a number of patches similar to it across the image \cite{xu2015patch}. The nonlocal transform domain framework \cite{dabov2007image} incorporates the NLSS prior, sparse representation \cite{elad2006image} and transform domain techniques \cite{yaroslavsky2001transform, maggioni2012nonlocal} into a subtle paradigm, which mainly follows three consecutive steps: grouping, collaborative filtering and aggregation. Briefly, given a local reference patch $\mathcal{P}_{ref}$, the grouping step aims to search for its similar patches with certain patch matching criteria \cite{foi2007pointwise, ehret2017global} and stacks them into a group $\mathcal{G}_{n}$ of higher dimension. To utilize the nonlocal similarity feature and estimate clean underlying patches $\mathcal{G}_c$, collaborative filtering is then performed on the noisy patch group $\mathcal{G}_{n}$ with different transforms and regularization terms \cite{beck2009fast, pang2017graph} via
\begin{equation}\label{tensor_collaborative_filtering}
  \hat{\mathcal{G}}_c = \mathop{\arg\min_{\mathcal{G}_c}} \| \mathcal{G}_n - \mathcal{G}_c \|_{F}^2 + \rho\cdot\Psi(\mathcal{G}_c)
  \vspace{-6pt}
\end{equation}
where $\| \mathcal{G}_n - \mathcal{G}_c \|_{F}^2$  measures the conformity between the clean and noisy groups, and $\Psi(\cdot)$ is a regularization term for certain priors. Finally, the aggregation step averagely writes back estimated clean image patches to their original location to further smooth out noise. \\
\indent \textbf{Color image denoising.} To handle color images, various strategies and representation are utilized to model the strong inter-channel correlation and different noise statistics in RGB channels. Briefly, Dai \textit{et al.} \cite{dai2013multichannel} adopt a multichannel fusion scheme according to a penalty function. Xu \textit{et al.} \cite{xu2017multi} concatenate the patches of RGB channels as a long vector and introduces weight matrices to characterize the realistic noise property and the sparsity prior. To avoid vectorization of image patches and preserve more structural information, some resent works take advantage of tensor representation. For example, Chang \textit{et al.} \cite{chang2017hyper} exploit the low-rank tensor recovery model to capture the spatial and spectral correlation. Rajwade \textit{et al.} \cite{rajwade2012image} use a multiway filtering strategy and apply the higher-order SVD (HOSVD). Kong \textit{et al.} \cite{kong2019color} introduce a global t-SVD transform for feature extraction of local patches. Despite the effectiveness of tensor-based methods, the tensor representation does not guarantee steady improvements \cite{kong2023comparison}.\\
\indent \textbf{Green channel prior.} The human visual system (HVS) is sensitive to green color because its peak sensitivity lies in the medium wavelengths, corresponding to the green portion. Therefore, the Bayer pattern is adopted by many digital cameras \cite{li2008image}, and the greater sampling rate of the green image will lead to higher SNR in the green channel \cite{guo2021joint}. Fig. \ref{Fig_SNR_comparison} illustrates the SNR difference in RGB channels of images from the CC15 dataset \cite{nam2016holistic}. It can be seen that the green channel of almost all noisy images has a higher SNR than red/blue channels. To exploit the GCP, existing methods mainly focus on performing demosaicing and denoising on raw images. For example, GCP-Net \cite{guo2021joint} utilizes the green channel to guide the feature extraction and feature upsampling of the whole image. SGNet \cite{liu2020joint} produces an initial estimate of the green channel and then uses it as a guidance to recover all missing values.
\vspace{-6pt}
\vspace{-8pt}
\\
\indent To the best of our knowledge, the GCP has not been exploited to guide denoising in the sRGB space. In this paper, we intend to fill this gap and propose GCP-ID, a simple GCP-guided color image denosing method with block circulant representation and t-SVD transform. It is worth mentioning that t-SVD has been adopted by different denoisers \cite{zhang2014novel, kong2019color, shi2021robust}, but they all treat each channel as equal.
\vspace{-16pt}
\section{Method}
\begin{figure*}[htbp]
  \centering
  \graphicspath{{Figs/}}
  \includegraphics[width=6.899in, height=1.239in]{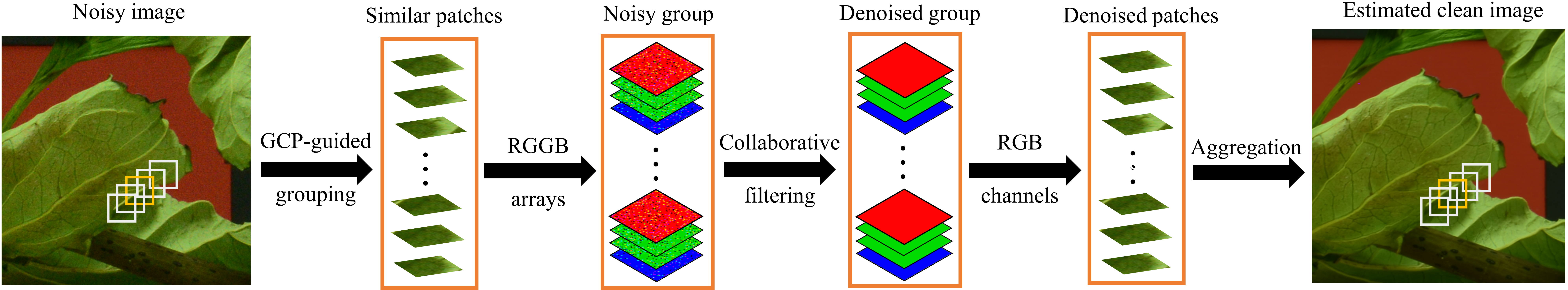}\\
  \caption{Flowchart of the proposed GCP-ID method for sRGB color image denoising.}
  \vspace{-15pt}
  \label{Figs_flowchart}
\end{figure*}

In this section, we present our GCP-ID method in detail. Briefly, GCP-ID consists of three key ingredients: 1) green channel guided patch search, 2) RGGB representation for image patches and 3) nonlocal t-SVD transform. The overall denoising procedure is illustrated in Fig. \ref{Figs_flowchart}.



\subsection{Green Channel Guided Patch Search}\label{section_method_patch_search}
Given a color image patch $\mathcal{P}_{ref} \in \mathbb{R}^{N \times N \times 3}$, our goal is to search its $K$ most similar patches in a local window by Euclidean distance. Directly comparing image patches will incur high computational burden. Besides, the presence of complex real-world noise may undermine the patch search process. To efficiently group more closely similar patches, we consider to take advantage of the green channel, which is less noisier and share structural information with red/blue channels. More specifically, we calculate the distance $d$ between image patches via
\begin{equation}\label{Equ_distance_calculation}
  d=\left\{
    \begin{aligned}
    &\| \mathcal{P}_{ref}^G -  \mathcal{P}_j^G \|, \, \|\mathcal{P}_{ref}^G\| \geq \text{max}(\lambda \|\mathcal{P}_{ref}^R\|,\lambda \|\mathcal{P}_{ref}^B\|) \\
    & \| \mathcal{P}_{ref}^{mean} -  \mathcal{P}_j^{mean} \|_F, \, \text{otherwise}
    \end{aligned}
    \right.
 \end{equation}
where $\mathcal{P}^{R}$, $\mathcal{P}^{G}$ and $\mathcal{P}^{B}$ represent R, G and B channels, respectively. $\mathcal{P}^{mean}$ is the mean value of RGB channels, which can be regarded as the opponent color space or average pooling operation on RGB channels. $\lambda$ is a threshold parameter that controls the search scheme. In our paper, we empirically set $\lambda = 0.8$, and Eq. (\ref{Equ_distance_calculation}) indicates that the green channel is used for local patch search if its importance is considered larger than certain weights. \\
\indent To verify the effectiveness of the green channel guided strategy, we compare the patch search successful rate measured by the ratio of patches similar in the underlying clean image. Specifically, for each noisy input of CC15 dataset, we randomly select 1000 reference patches $\mathcal{P}_{ref}\in \mathbb{R}^{8\times8\times3}$ and search 60 similar patches for $\mathcal{P}_{ref}$ in a local window of size $20\times20$. Table \ref{Table_patch_search_successful_rate} reports the average successful rate of different search schemes. The advantage of the green guided strategy will help encourage sparsity in the transform domain, and the correlation among similar image patches can be better captured by learning group-level transform $\mathbf{U}_{group} \in \mathbb{R}^{K\times K}$ via classic methods such as the principal component analysis (PCA) \cite{dabov2009bm3d}.
\begin{table}[htbp]
\small
  \centering
  \caption{Average patch search successful rate of different schemes for noisy images from the CC15 dataset. `Green only': search using only the green channel. `Opp': Opponent color space.}
  \scalebox{0.938}{
    \begin{tabular}{ccccc}
    \toprule
    Search scheme & Green only & YCbCr & Opp  & Green guided \\
    \midrule
    Successful rate & 0.4881 & 0.4992 & 0.4938 & \textbf{0.5035} \\
    \bottomrule
    \end{tabular}
    }%
  \label{Table_patch_search_successful_rate}%
\end{table}%
\vspace{-6pt}

\subsection{RGGB Representation}
\vspace{-3pt}
\begin{figure}[htbp]
\graphicspath{{Figs/}}
  \centering
  \includegraphics[width=3.099in, height=0.659in]{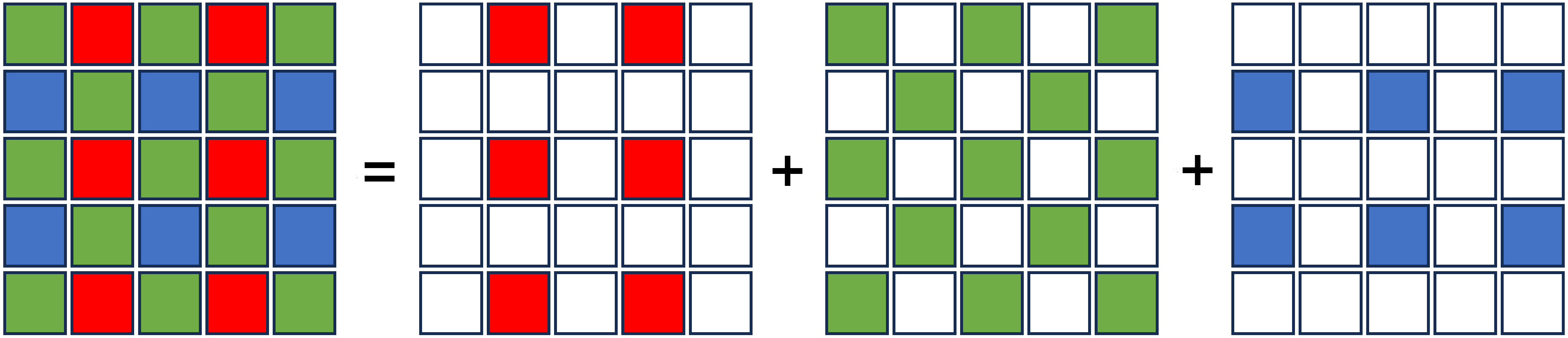}\\
  \caption{Illustration of the Bayer CFA pattern.}
  \label{Fig_Bayer_pattern}
\end{figure}

\begin{figure}[htbp]
\graphicspath{{Figs/}}
  \centering
  \includegraphics[width=2.999in, height=0.869in]{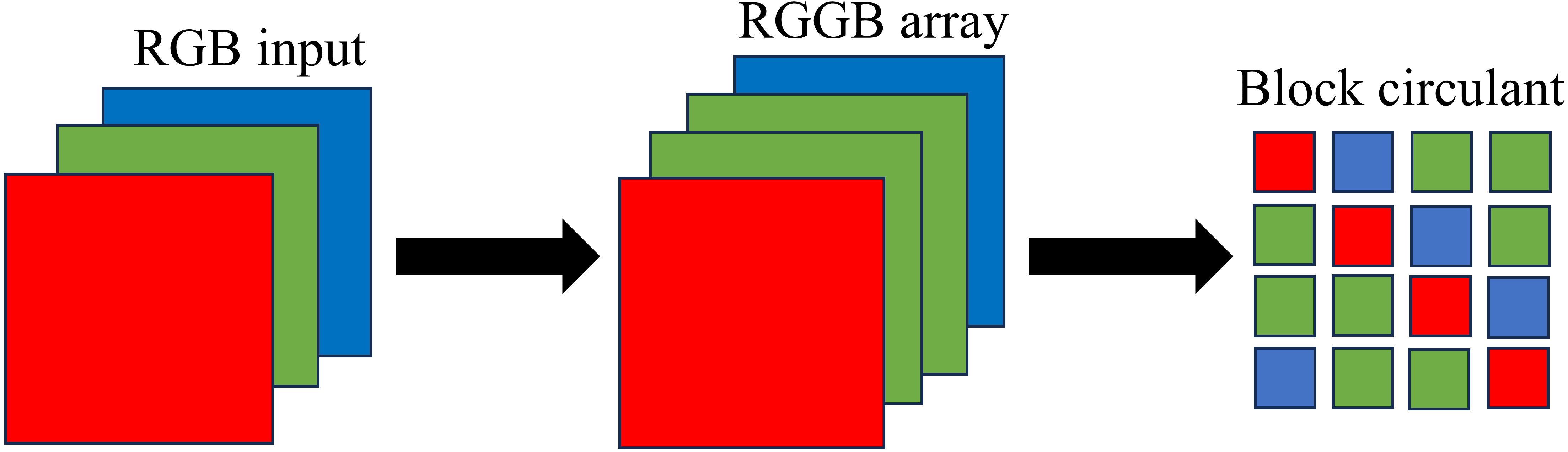}\\
  \caption{Block circulant representation for an RGB input.}
  \label{Figs_RGGB_representation}
\end{figure}

\vspace{-3pt}
As shown in Fig. \ref{Fig_Bayer_pattern}, the popular Bayer CFA pattern measures the green image on a quincunx grid and the red and blue images on rectangular grids\cite{li2008image}. As a result, the density of the green samples is twice than that of the other two. In practice, the raw Bayer pattern data is not always available, and the interpolation algorithm to reconstruct a full color representation of the image varies according to different camera devices.\\
\indent To explicitly model the density and importance of the green channel, we reshape each color image patch $\mathcal{P}\in \mathbb{R}^{ps\times ps \times 3}$ into an RGGB array $\hat{\mathcal{P}}\in \mathbb{R}^{ps\times ps \times 4}$ by inserting one extra green channel. Furthermore, to capture the spectral correlation among RGB channels, we use the block circulant matrix $\hat{\mathbf{P}}_{bcirc}\in \mathbb{R}^{4ps\times4ps}$ to represent the RGGB array. Fig. \ref{Figs_RGGB_representation} illustrates the RGGB and block circulant representation (BCR) for an RGB input. The BCR indicates that the density of green channels remains twice than that of red/blue ones in the sRGB space.
\vspace{-3pt}

\vspace{-10pt}
\subsection{Nonlocal t-SVD transform}
The stacked similar patches often share the same feature space, and a nonlocal generalization of SVD \cite{rajwade2012image} can be explored by learning pairwise projection matrices $\mathbf{U}_{row} \in \mathbb{R}^{4ps\times4ps}$ and $\mathbf{U}_{col}\in \mathbb{R}^{4ps\times4ps}$ via
\begin{equation}\label{Equ_nonlocal_SVD}
\begin{aligned}
  & \text{min} \sum_{i = 1}^K \| \hat{\mathbf{P}}_{bcirc_i} -  \mathbf{U}_{row} \hat{\mathbf{S}}_{bcirc_i} \mathbf{U}_{col}^T\|^2 \\
  & \text{s.t} \quad \mathbf{U}_{row}^T \mathbf{U}_{row} = \mathbf{I}, \, \mathbf{U}_{col}^T \mathbf{U}_{col} = \mathbf{I}
\end{aligned}
\end{equation}
where $\hat{\mathbf{S}}_{bcirc_i} \in \mathbb{R}^{4ps\times4ps}$ is the coefficient matrix of $\hat{\mathbf{P}}_{bcirc_i}$. From Fig. \ref{Figs_RGGB_representation}, it is noticed that the RGB channels of $\hat{\mathbf{P}}_{bcirc}$ repeat several times. To efficiently leverage such patch-level redundancy and exploit the spectral correlation among RGB channels, we can take advantage of the nonlocal t-SVD transform to simplify Eq. (\ref{Equ_nonlocal_SVD}) via
\vspace{-5.8pt}
\begin{equation}\label{Equ_nonlocal_t-SVD}
\begin{aligned}
  & \text{min} \sum_{i = 1}^K \| \hat{\mathcal{P}}_{i} -  \mathcal{U}_{row} * \hat{\mathcal{S}}_{i} * \mathcal{U}_{col}^T\|_F^2 \\
  & \text{s.t} \quad \mathcal{U}_{row}^T * \mathcal{U}_{row} = \mathcal{I}, \, \, \mathcal{U}_{col}^T * \mathcal{U}_{col} = \mathcal{I}
\end{aligned}
\end{equation}
where $\mathcal{U}_{row} \in \mathbb{R}^{ps\times ps \times 4}$ and $\mathcal{U}_{col}\in \mathbb{R}^{ps\times ps \times 4}$ are a pair of orthogonal tensors,  $\hat{\mathcal{S}}_{i} \in \mathbb{R}^{ps\times ps \times 4}$ is the corresponding coefficient tensor of $\hat{\mathcal{P}}_{i}$.
Eq. (\ref{Equ_nonlocal_t-SVD}) can be solved by slicewise SVD in the Fourier domain after performing the FFT along the third mode of $\hat{\mathcal{P}}$ via
\begin{equation}\label{Equ_FFT}
\begin{aligned}
\hat{\mathcal{P}} \times _3\mathbf{W}_{FFT} & = \begin{pmatrix}
1 & 1 & 1 & 1\\
1 & -i & -1 & i \\
1 & -1 & 1 & -1 \\
1 & i & -1 & -i \\
\end{pmatrix}
\begin{pmatrix}
R \\
G \\
G \\
B \\
\end{pmatrix}\\
& = \begin{pmatrix}
R + 2G + B \\
R-G + (B-G)i \\
R - B \\
R-G + (G-B)i \\
\end{pmatrix}
\end{aligned}
\end{equation}
Interestingly, from Eq. (\ref{Equ_FFT}) it can be seen that in the Fourier domain, the first slice models the density of the green channel, the second and fourth slices are complex conjugate that exploit the relationship among RGB channels, and the third slice captures the difference between red and blue channels. Therefore, the RGGB representation also captures the correlation among RGB channels.\\
\indent Given a group of noisy BCR patches $\hat{\mathcal{G}}_n \in \mathbb{R}^{ps\times ps \times 4 \times K}$, the overall denoising process of the proposed GCP-ID can be summarized using a one step extension of t-SVD
\begin{equation}\label{Equ_summarize_denoising}
  \hat{\mathcal{S}}_n = (\mathcal{U}_{row}^T * \hat{\mathcal{G}}_n * \mathcal{U}_{col}) \times _4 \mathbf{U}_{group}
\end{equation}
The hard-thresholding technique \cite{donoho1994ideal} can be applied to shrink the coefficient $\hat{\mathcal{S}}_n$ under a certain threshold $\tau$ via
 \begin{equation}\label{Equ_hard_thresholding}
  \hat{\mathcal{S}}_{truncate}=\left\{
    \begin{aligned}
    \hat{\mathcal{S}}_n, \quad |\hat{\mathcal{S}}_n| \geq \tau \\
    0, \quad |\hat{\mathcal{S}}_n| < \tau
    \end{aligned}
    \right.
 \end{equation}
The estimated clean group $\hat{\mathcal{G}}_c$ can be reconstructed by the inverse transform of Eq. (\ref{Equ_summarize_denoising}) via
\begin{equation}\label{Equ_summarize_inverse_denoising}
\hat{\mathcal{G}}_c = (\mathcal{U}_{row} * \hat{\mathcal{S}}_{truncate} * \mathcal{U}_{col}^T) \times _4 \mathbf{U}_{group}^T
\end{equation}
Finally, we fetch the RGB channels of $\hat{\mathcal{G}}_c$, and each denoised image patch $\mathcal{P}_c \in \mathbb{R}^{ps\times ps \times 3}$ is averagely written back to its original location. The implementation of GCP-ID is briefed in Algorithm \ref{Algorithm_GCP-ID}.
\setlength{\textfloatsep}{8pt}
\begin{algorithm}[ht]
\caption{GCP-ID}
{\bf Input:} Noisy image $\mathcal{Y} \in \mathbb{R}^{M \times N \times 3}$, patch size $ps$, number of similar patches $K$ and search window size $W$.\\
{\bf Output:} Estimated clean image $\mathcal{X} \in \mathbb{R}^{M \times N \times 3}$.\\
{\bf Step 1} (Grouping): For each reference patch $\mathcal{P}\in \mathbb{R}^{ps \times ps \times 3}$ of $\mathcal{Y}$, search its $K$ most similar patches within $W$ according to Eq. (\ref{Equ_distance_calculation}) and stack them in a group $\hat{\mathcal{G}}_n \in \mathbb{R}^{ps\times ps \times 4 \times K}$ based on RGGB representation.\\
{\bf Step 2} (Collaborative filtering):\\
 \hspace*{0.18in}(1) Perform the forward nonlocal t-SVD based on Eq. (\ref{Equ_summarize_denoising}) to obtain the coefficient tensor $\hat{\mathcal{S}}_n$, pairwise projection tensors $\mathcal{U}_{row}$, $\mathcal{V}_{row}$ and group-level transform $\mathbf{G}_{group}^T$.\\
 \hspace*{0.18in}(2) Apply the hard-threshold technique to $\hat{\mathcal{S}}_n$ to obtain the truncated coefficient tensor $\hat{\mathcal{S}}_{truncate}$.\\
 \hspace*{0.18in}(3) Take the inverse nonlocal t-SVD transform in Eq. (\ref{Equ_summarize_inverse_denoising}) to obtain estimated clean group $\hat{\mathcal{G}}_c$.\\
{\bf Step 3} (Aggregation): Fetch the RGB channels of $\hat{\mathcal{G}}_c$ and averagely write back all denoised image patches.
\label{Algorithm_GCP-ID}
\end{algorithm}

\subsection{Complexity Analysis}
For each reference patch of noisy observation $\mathcal{Y}$, the computational burden of GCP-ID lies mainly in three parts: 1) the search of $K$ similar patches $O(KW^2ps)$, 2) the group-level PCA transform $O(Kps^4)$ and 3) the nonlocal t-SVD transform $O(Kps^3)$. The overall complexity of GCP-ID is $O([KW^2ps + Kps^4]n_{ref})$, where $n_{ref}$ is the number of reference patches of $\mathcal{Y}$.
\vspace{-1pt}
\section{Experiments}
In this section, we mainly report the results of for real-world color image and video denoising tasks. All implementations and source code of related methods are provided by the authors or downloaded from the authors personal websites. We employ peak signal-to-noise ratio (PSNR) and structure similarity (SSIM \cite{wang2004image}) for performance evaluations. Normally, the larger these two measures are, the closer the denoised image is to the reference one.
\subsection{Real-world color image denoising}
\indent \textbf{Datasets.} We perform experiments and comparison on popular benchmark datasets of real-world color image denoising, \textit{i.e.}, SIDD \cite{abdelhamed2018high}, DND \cite{plotz2017benchmarking}, CC15 \cite{nam2016holistic}, CC60 \cite{nam2016holistic}, PolyU \cite{xu2018real}, HighISO \cite{yue2019high}, and IOCI \cite{kong2023comparison}, where SIDD provides training/validation samples. The clean-noisy image pairs of DnD and SIDD are not available, and the test results can be acquired by submitting denoised images.\\
\indent \textbf{Implementation details.} GCP-ID has several decisive parameters: the patch size $ps$, search window size $W$, number of similar patches $K$, hard threshold value $\tau$ and noise level $\sigma$. In our experiments, we set $ps = 8$, $W = 20$, $K = 30$, $\tau = 1.1\sigma\sqrt{2\text{log}(3ps^2K)}$. Besides, $\sigma$ controls the sparsity of coefficients and it varies according to different datasets. Specifically, for SIDD, $\sigma$ is chosen from $[10,150]$ based on the validation set. For DND, $\sigma$ is decided by the optimal test result of four submissions. For other datasets, $\sigma$ is chosen from $[10, 40]$ with a stepsize of $10$.\\
\indent \textbf{Compared methods.} We compare GCP-ID with both traditional denoisers and DNN methods. Briefly, traditional denoisers include state-of-the-art nonlocal approaches such as CBM3D \cite{dabov2007color}, CMSt-SVD \cite{kong2019color}, TWSC \cite{xu2018trilateral} and NLHCC \cite{hou2020nlh}. DNN methods consist of effective supervised (S) and self-supervised (SS) models such as FFDNet \cite{zhang2018ffdnet}, Restormer \cite{zamir2022restormer}, Self2Self \cite{quan2020self2self} and Neigh2Neigh \cite{huang2021neighbor2neighbor}. For a fair comparison, parameters and models of all compared methods are carefully chosen to obtain their best possible performance for each dataset \cite{kong2023comparison}.\\
\indent \textbf{Performance evaluation.} Table \ref{Table_real_world_color_image_denoising} lists the average PSNR and SSIM values of compared methods. Overall, the supervised DNN methods show outstanding performance on SIDD and DND datasets, but the pretrained models may not generalize well in the absence of the training/validation process. From Table \ref{Table_real_world_color_image_denoising}, it can be seen that the proposed GCP-ID is able to produce very competitive performance for the real-world color image denoising task. For example, compared to the classic nonlocal denoiser CBM3D and the representative self-supervised method Self2Self, GCP-ID provides the PSNR improvements of more than 0.3dB on the SIDD and CC datasets. In addition, the advantage of GCP-ID over CMSt-SVD on a variety of datasets and cameras demonstrates the effectiveness of GCP guided patch search strategy and RGGB patch representation.
\begin{table*}[htbp]
  \centering
  \caption{Average PSNR, SSIM values on real-world color image datasets. The best results are black bolded.}
  \renewcommand{\arraystretch}{0.1998}
  \scalebox{0.519}{
    \begin{tabular}{ccccccccccccccccc}
    \toprule
    \multicolumn{2}{c}{\multirow{2}[4]{*}{Methods}} & \multirow{2}[4]{*}{DND} & \multirow{2}[4]{*}{SIDD} & \multirow{2}[4]{*}{CC15} & \multirow{2}[4]{*}{CC60} & \multirow{2}[4]{*}{PolyU} & \multirow{2}[4]{*}{HighISO} & \multicolumn{9}{c}{IOCI} \\
\cmidrule{9-17}    \multicolumn{2}{c}{} &       &       &       &       &       &       & CAN 600D & CAN 5DMark4 & Fuji X100T & HW honor6X & IPHONE13 & NIK D5300 & OPPO R11s & SONY A6500 & XIAOMI8 \\
    \midrule
    -     & \# Images & 1000  & 1280  & 15    & 60    & 100   & 100   & 25    & 83    & 71    & 30    & 174   & 56    & 39    & 36    & 50 \\
    \midrule
    \multirow{18}[36]{*}{\shortstack[c]{Traditional \\ denoisers}} & \multirow{2}[4]{*}{NLHCC \cite{hou2020nlh}} & 38.85  & -     & \textbf{38.49} & 39.86  & 38.36  & 40.29  & 42.72  & 43.66  & 42.80  & 38.84  & 41.13  & 43.25  & -     & \textbf{46.02} & 35.73  \\
\cmidrule{3-17}          &       & 0.953  & -     & \textbf{0.965} & 0.976  & 0.965  & 0.971  & 0.984  & 0.986  & 0.978  & 0.959  & 0.976  & 0.981  & -     & \textbf{0.991} & 0.955  \\
\cmidrule{2-17}          & \multirow{2}[4]{*}{Bitonic\cite{treece2022real}} & 37.85  & 36.67  & 35.22  & 35.98  & 36.64  & 37.37  & 39.63  & 40.76  & 41.05  & 37.71  & 39.09  & 39.22  & 38.87  & 43.25  & 34.92  \\
\cmidrule{3-17}          &       & 0.936  & 0.933  & 0.925  & 0.931  & 0.940  & 0.943  & 0.952  & 0.965  & 0.964  & 0.940  & 0.952  & 0.954  & 0.959  & 0.979  & 0.941  \\
\cmidrule{2-17}          & \multirow{2}[4]{*}{LLRT \cite{chang2017hyper}} & 35.45  & 30.74  & 37.77  & 39.76  & 38.28  & 39.59  & 42.24  & 42.68  & 42.22  & 37.91  & 42.02  & 41.76  & 38.69  & 45.17  & 35.71  \\
\cmidrule{3-17}          &       & 0.897  & 0.766  & 0.957  & 0.977  & 0.970  & 0.972  & 0.983  & 0.992  & 0.975  & 0.969  & 0.984  & 0.979  & 0.972  & 0.989  & 0.962  \\
\cmidrule{2-17}          & \multirow{2}[4]{*}{MCWNNM \cite{xu2017multi}} & 37.38  & 29.54  & 37.02  & 38.54  & 38.26  & 39.89  & 42.07  & 44.22  & 42.48  & 39.46  & 41.33  & 41.74  & 40.71  & 45.38  & 35.84  \\
\cmidrule{3-17}          &       & 0.929  & 0.888  & 0.950  & 0.967  & 0.965  & 0.970  & 0.979  & 0.988  & 0.976  & 0.961  & 0.976  & 0.975  & 0.973  & 0.990  & 0.952  \\
\cmidrule{2-17}          & \multirow{2}[4]{*}{TWSC \cite{xu2018trilateral}} & 37.96  & -     & 37.90  & 39.66  & 38.62  & \textbf{40.62} & 42.52  & \textbf{44.94} & 42.26  & 38.71  & 41.71  & 42.23  & 40.65  & 45.49  & 35.40  \\
\cmidrule{3-17}          &       & 0.942  & -     & 0.959  & 0.976  & 0.967  & \textbf{0.975} & 0.982  & \textbf{0.992} & 0.973  & 0.945  & 0.980  & 0.975  & 0.972  & 0.990  & 0.939  \\
\cmidrule{2-17}          & \multirow{2}[4]{*}{CBM3D \cite{dabov2007color}} & 37.73  & 34.74  & 37.69  & 39.41  & 38.69  & 40.35  & 42.54  & 44.74  & 42.65  & 39.97  & 42.03  & 42.20  & 40.75  & 45.72  & 36.38  \\
\cmidrule{3-17}          &       & 0.934  & 0.922  & 0.957  & 0.975  & 0.970  & 0.974  & 0.984  & 0.992  & 0.977  & 0.967  & 0.983  & 0.979  & 0.974  & 0.990  & 0.961  \\
\cmidrule{2-17}          & \multirow{2}[4]{*}{4DHOSVD \cite{rajwade2012image}} & 37.58  & 34.49  & 37.52  & 39.15  & 38.54  & 40.27  & 42.19  & 44.49  & 42.60  & 39.82  & 41.75  & 41.82  & 40.71  & 45.58  & 36.27  \\
\cmidrule{3-17}          &       & 0.929  & 0.911  & 0.956  & 0.973  & 0.968  & 0.973  & 0.982  & 0.990  & 0.976  & 0.966  & 0.980  & 0.977  & 0.974  & 0.990  & 0.961  \\
\cmidrule{2-17}          & \multirow{2}[4]{*}{CMSt-SVD \cite{kong2019color}} & 38.25  & 34.38  & 37.95  & 39.76  & 38.85  & 40.49  & 42.75  & 44.65  & 42.68  & 40.08  & \textbf{42.06} & 42.72  & 40.87  & 45.91  & 36.40  \\
\cmidrule{3-17}          &       & 0.940  & 0.900  & 0.959  & 0.976  & 0.971  & 0.974  & 0.984  & 0.991  & 0.977  & 0.967  & \textbf{0.982} & 0.980  & 0.974  & 0.991  & 0.962  \\
\cmidrule{2-17}          & \multirow{2}[4]{*}{GCP-ID} & 38.38  & 35.00  & 38.30  & \textbf{40.15} & \textbf{38.90} & 40.56  & \textbf{43.02} & 44.60  & \textbf{42.70} & \textbf{40.18} & \textbf{42.05}  & \textbf{42.98} & \textbf{40.96} & \textbf{46.06} & \textbf{36.48} \\
\cmidrule{3-17}          &       & 0.941  & 0.915  & 0.962  & \textbf{0.978} & \textbf{0.971} & 0.974  & \textbf{0.985} & 0.991  & \textbf{0.977} & \textbf{0.968} & \textbf{0.982}  & \textbf{0.981} & \textbf{0.975} & \textbf{0.991} & \textbf{0.962} \\
    \midrule
    \midrule
    \multirow{10}[20]{*}{\shortstack[c]{DNN (SS)}} & \multirow{2}[4]{*}{AP-BSN \cite{lee2022ap}} & 37.29  & 35.97  & 35.44  & 36.76  & 36.99  & 38.26  & 41.29  & 42.72  & 41.40  & 37.87  & 40.83  & 40.86  & 39.55  & 43.73  & 34.37  \\
\cmidrule{3-17}          &       & 0.932  & 0.925  & 0.936  & 0.956  & 0.956  & 0.965  & 0.979  & 0.986  & 0.970  & 0.949  & 0.977  & 0.973  & 0.969  & 0.983  & 0.937  \\
\cmidrule{2-17}          & \multirow{2}[4]{*}{C2N \cite{jang2021c2n}} & 37.28  & -     & 37.02  & -     & 37.69  & 38.86  & 41.95  & 42.61  & 41.33  & 38.73  & 40.50  & 40.78  & 40.05  & 44.64  & 35.53  \\
\cmidrule{3-17}          &       & 0.924  & -     & 0.945  & -     & 0.958  & 0.960  & 0.980  & 0.980  & 0.966  & 0.954  & 0.977  & 0.964  & 0.970  & 0.987  & 0.952  \\
\cmidrule{2-17}          & \multirow{2}[4]{*}{Neigh2Neigh \cite{huang2021neighbor2neighbor}} & -     & -     & 34.47  & 35.65  & 37.10  & 36.59  & 40.41  & 41.29  & 41.24  & 38.13  & 39.63  & 38.05  & 39.17  & 44.49  & 35.30  \\
\cmidrule{3-17}          &       & -     & -     & 0.883  & 0.907  & 0.942  & 0.908  & 0.960  & 0.966  & 0.958  & 0.940  & 0.952  & 0.919  & 0.958  & 0.984  & 0.944  \\
\cmidrule{2-17}          & \multirow{2}[4]{*}{SASL \cite{li2023spatially}} & 38.00  & -     & 34.93  & -     & 37.13  & 38.24  & 41.86  & 43.07  & 40.29  & 37.22  & 40.85  & 41.64  & 39.10  & 42.89  & 34.25  \\
\cmidrule{3-17}          &       & 0.9364  & -     & 0.9356  & -     & 0.9540  & 0.9641  & 0.9815  & 0.9868  & 0.9632  & 0.9473  & 0.9768  & 0.9762  & 0.9650  & 0.9796  & 0.9298  \\
\cmidrule{2-17}          & \multirow{2}[4]{*}{Self2Self\cite{quan2020self2self}} & -     & -     & 36.26  & 38.23  & -     & 39.49  & -     & -     & -     & -     & -     & -     & -     & -     & - \\
\cmidrule{3-17}          &       & -     & -     & 0.947  & 0.969  & -     & 0.963  & -     & -     & -     & -     & -     & -     & -     & -     & - \\
    \midrule
    \midrule
    \multirow{24}[48]{*}{\shortstack[c]{DNN (S)}} & \multirow{2}[4]{*}{AINDNet\cite{Kim_2020_CVPR}} & 39.77  & 39.08  & 36.14  & 37.19  & 37.33  & 38.00  & 39.33  & 39.49  & 38.50  & 36.53  & 37.27  & 38.11  & 37.44  & 40.17  & 34.65  \\
\cmidrule{3-17}          &       & 0.959  & 0.953  & 0.935  & 0.949  & 0.954  & 0.946  & 0.976  & 0.979  & 0.966  & 0.954  & 0.964  & 0.961  & 0.968  & 0.981  & 0.953  \\
\cmidrule{2-17}          & \multirow{2}[4]{*}{CBDNet\cite{guo2019toward}} & 38.06  & 33.26  & 36.20  & 37.67  & 37.81  & 38.18  & 42.41  & 42.55  & 41.88  & 38.35  & 40.63  & 40.92  & 39.54  & 44.38  & 35.54  \\
\cmidrule{3-17}          &       & 0.942  & 0.869  & 0.919  & 0.940  & 0.956  & 0.942  & 0.981  & 0.980  & 0.971  & 0.946  & 0.968  & 0.963  & 0.965  & 0.981  & 0.950  \\
\cmidrule{2-17}          & \multirow{2}[4]{*}{CycleISP \cite{zamir2020cycleisp}} & 39.57  & 39.42  & 35.40  & 36.87  & 37.61  & 37.70  & 41.84  & 42.64  & 41.59  & 38.47  & 40.61  & 40.03  & 39.86  & 44.75  & 35.54  \\
\cmidrule{3-17}          &       & 0.955  & 0.956  & 0.916  & 0.939  & 0.955  & 0.936  & 0.977  & 0.978  & 0.969  & 0.952  & 0.970  & 0.953  & 0.968  & 0.987  & 0.949  \\
\cmidrule{2-17}          & \multirow{2}[4]{*}{DeamNet\cite{ren2021adaptive}} & 39.63  & 39.35  & 36.63  & -     & 37.70  & 36.93  & 40.86  & 41.22  & 39.72  & 33.67  & 40.25  & 38.92  & 39.56  & 40.52  & 34.61  \\
\cmidrule{3-17}          &       & 0.953  & 0.955  & 0.936  & -     & 0.958  & 0.944  & 0.978  & 0.979  & 0.967  & 0.953  & 0.966  & 0.957  & 0.969  & 0.980  & 0.952  \\
\cmidrule{2-17}          & \multirow{2}[4]{*}{DIDN\cite{yu2019deep}} & 39.64  & \textbf{39.78} & 36.06  & -     & 37.36  & 38.24  & 41.68  & 42.35  & 41.17  & 38.15  & 40.03  & 40.32  & 39.73  & 44.07  & 35.28  \\
\cmidrule{3-17}          &       & 0.953  & \textbf{0.958} & 0.946  & -     & 0.953  & 0.950  & 0.977  & 0.975  & 0.966  & 0.951  & 0.963  & 0.959  & 0.966  & 0.985  & 0.949  \\
\cmidrule{2-17}          & \multirow{2}[4]{*}{DnCNN\cite{zhang2017beyond}} & 37.90  & 37.73  & 37.47  & 39.32  & 38.51  & 40.01  & 41.91  & 44.16  & 41.57  & 39.92  & 42.03  & 42.12  & 40.26  & 43.97  & 36.30  \\
\cmidrule{3-17}          &       & 0.943  & 0.941  & 0.954  & 0.974  & 0.966  & 0.971  & 0.979  & 0.991  & 0.969  & 0.962  & 0.983  & 0.977  & 0.970  & 0.984  & 0.960  \\
\cmidrule{2-17}          & \multirow{2}[4]{*}{FFDNet\cite{zhang2018ffdnet}} & 37.61  & 38.27  & 37.67  & 39.73  & 38.76  & 40.28  & 42.55  & 44.77  & 42.44  & 40.05  & 42.43  & 42.44  & 40.75  & 45.71  & 36.47  \\
\cmidrule{3-17}          &       & 0.942  & 0.948  & 0.956  & 0.977  & 0.970  & 0.973  & 0.982  & 0.992  & 0.976  & 0.967  & 0.984  & 0.979  & 0.973  & 0.990  & 0.961  \\
\cmidrule{2-17}          & \multirow{2}[4]{*}{InvDN\cite{liu2021invertible}} & 39.57  & 39.28  & 34.55  & 36.33  & 35.94  & 38.01  & 40.09  & 42.03  & 40.02  & 33.32  & 40.06  & 40.98  & 39.10  & 40.74  & 34.15  \\
\cmidrule{3-17}          &       & 0.952  & 0.955  & 0.937  & 0.953  & 0.947  & 0.952  & 0.973  & 0.978  & 0.960  & 0.930  & 0.966  & 0.971  & 0.964  & 0.969  & 0.935  \\
\cmidrule{2-17}          & \multirow{2}[4]{*}{MIRNet\cite{Zamir2020MIRNet}} & 39.88  & 39.55  & 36.06  & 37.25  & 37.49  & 38.10  & 41.68  & 42.76  & 41.05  & 38.28  & 40.28  & 40.46  & 39.73  & 43.66  & 35.42  \\
\cmidrule{3-17}          &       & 0.956  & 0.957  & 0.942  & 0.956  & 0.956  & 0.952  & 0.978  & 0.980  & 0.968  & 0.954  & 0.966  & 0.964  & 0.967  & 0.983  & 0.953  \\
\cmidrule{2-17}          & \multirow{2}[4]{*}{Restormer\cite{zamir2022restormer}} & \textbf{40.03} & \textbf{39.79} & 36.33  & -     & 37.66  & 38.29  & 41.84  & 42.49  & 41.47  & 38.42  & 40.13  & 40.53  & 39.56  & 44.19  & 35.65  \\
\cmidrule{3-17}          &       & \textbf{0.956} & \textbf{0.959} & 0.941  & -     & 0.956  & 0.948  & 0.979  & 0.976  & 0.968  & 0.952  & 0.963  & 0.962  & 0.963  & 0.986  & 0.953  \\
\cmidrule{2-17}          & \multirow{2}[4]{*}{RIDNet\cite{anwar2019real}} & 39.26  & 38.70  & 36.84  & 38.11  & 38.57  & 38.60  & 42.15  & 42.39  & 41.59  & 38.88  & 40.91  & 41.01  & 40.07  & 44.40  & 35.76  \\
\cmidrule{3-17}          &       & 0.953  & 0.950  & 0.941  & 0.961  & 0.970  & 0.956  & 0.980  & 0.989  & 0.975  & 0.962  & 0.978  & 0.977  & 0.972  & 0.987  & 0.957  \\
\cmidrule{2-17}          & \multirow{2}[4]{*}{VDIR\cite{soh2022variational}} & -     & 39.26  & 34.90  & 36.51  & 37.01  & 36.84  & 41.40  & 41.29  & 40.81  & 37.65  & 39.20  & 39.41  & 38.98  & 44.40  & 35.04  \\
\cmidrule{3-17}          &       & -     & 0.955  & 0.894  & 0.925  & 0.940  & 0.911  & 0.970  & 0.965  & 0.953  & 0.936  & 0.946  & 0.938  & 0.956  & 0.985  & 0.939  \\
    \bottomrule
    \end{tabular}}%
  \label{Table_real_world_color_image_denoising}%
  \vspace{-11.9pt}
\end{table*}%

\vspace{-13.9pt}

\begin{figure}[htbp]
\graphicspath{{Figs/CC15/combined/}}
\centering
\subfigure[Reference]{
\label{Fig4}
\includegraphics[width=1.03in, height=1.03in]{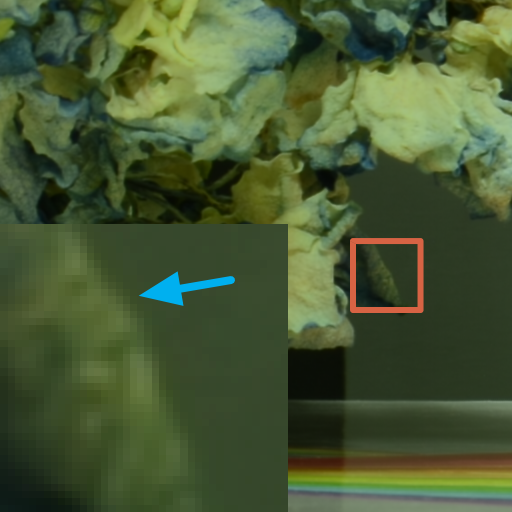}}
\subfigure[Noisy]{
\label{Fig4}
\includegraphics[width=1.03in, height=1.03in]{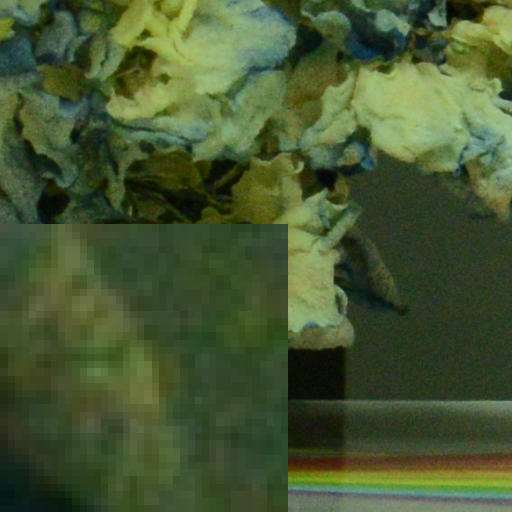}}
\subfigure[GCP-ID]{
\label{Fig4}
\includegraphics[width=1.03in, height=1.03in]{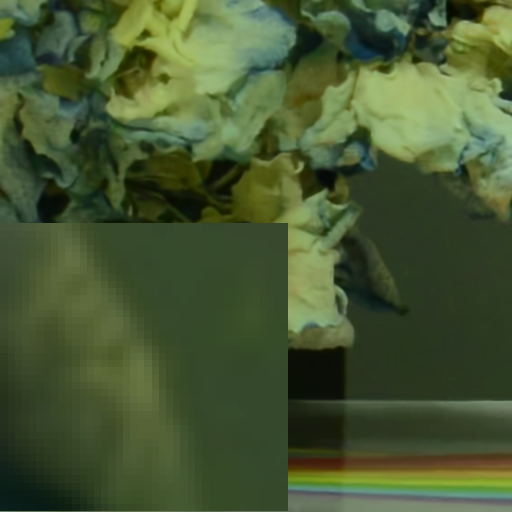}} \\
\vspace{-9.18pt}
\subfigure[SASL]{
\label{Fig4}
\includegraphics[width=1.03in, height=1.03in]{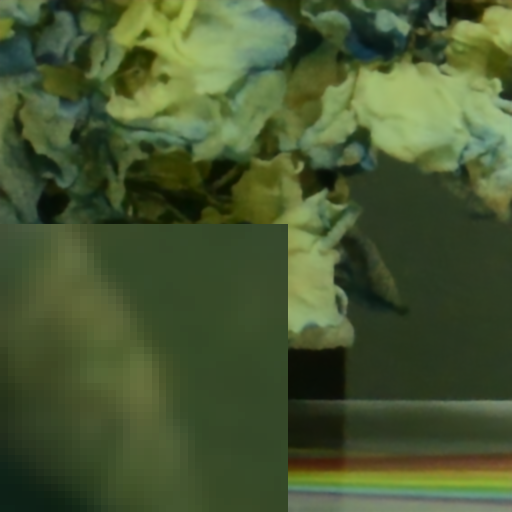}}
\subfigure[FFDNet]{
\label{Fig4}
\includegraphics[width=1.03in, height=1.03in]{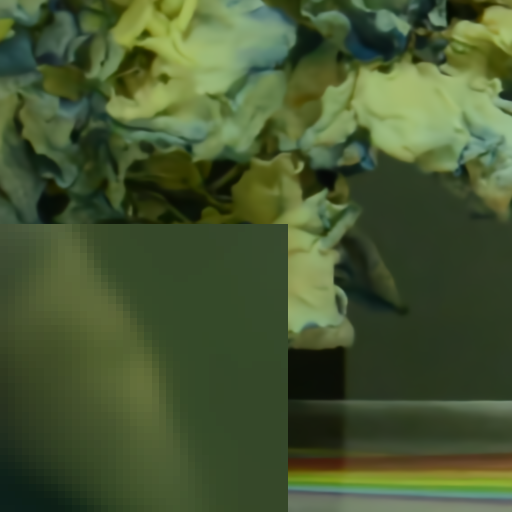}}
\subfigure[Restormer]{
\label{Fig4}
\includegraphics[width=1.03in, height=1.03in]{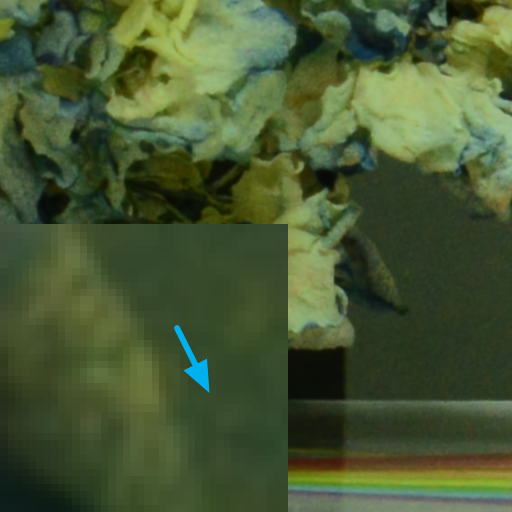}} \\
\vspace{-9.18pt}
\subfigure[NLHCC]{
\label{Fig4}
\includegraphics[width=1.03in, height=1.03in]{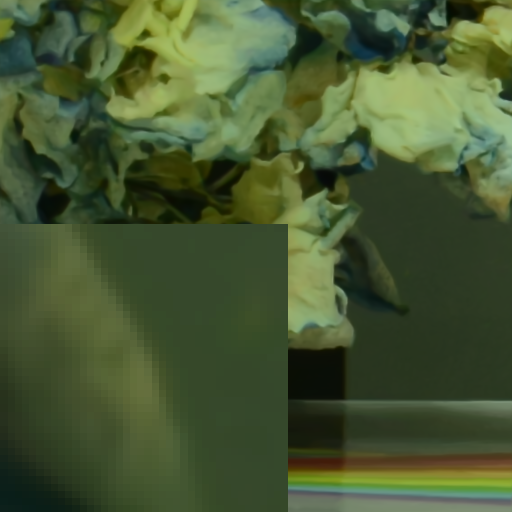}}
\subfigure[CBM3D]{
\label{Fig4}
\includegraphics[width=1.03in, height=1.03in]{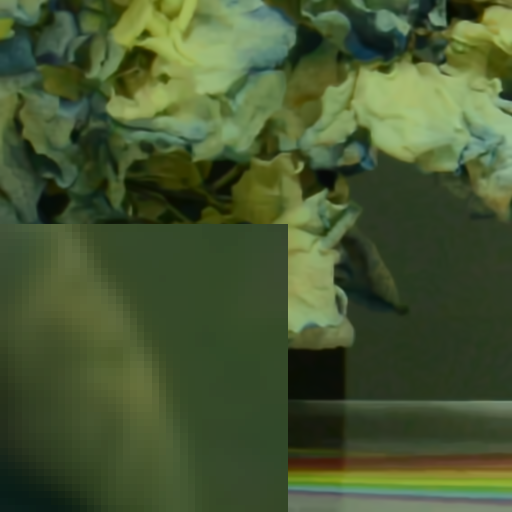}}
\subfigure[CMSt-SVD]{
\label{Fig4}
\includegraphics[width=1.03in, height=1.03in]{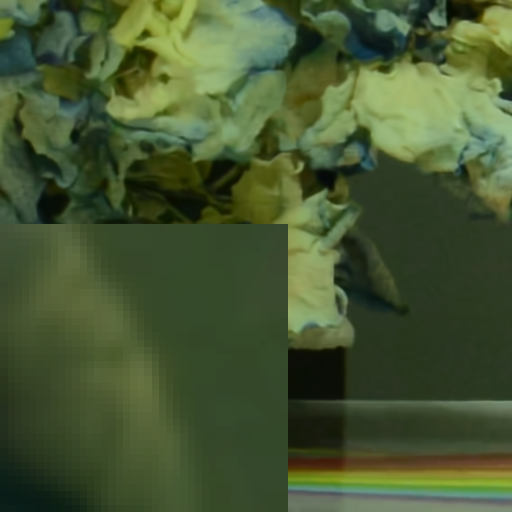}}

\caption{Color image denoising results on the CC15 dataset.}
\label{Fig_CC15_visual_evaluation}
\end{figure}

\begin{figure}[htbp]
\graphicspath{{Figs/DND/combined/}}
\centering
\subfigure[Noisy]{
\label{Fig4}
\includegraphics[width=0.758in, height = 0.758in]{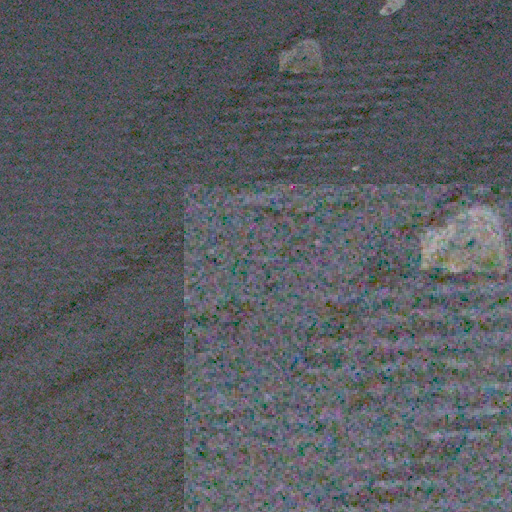}}
\subfigure[GCP-ID]{
\label{Fig4}
\includegraphics[width=0.758in, height = 0.758in]{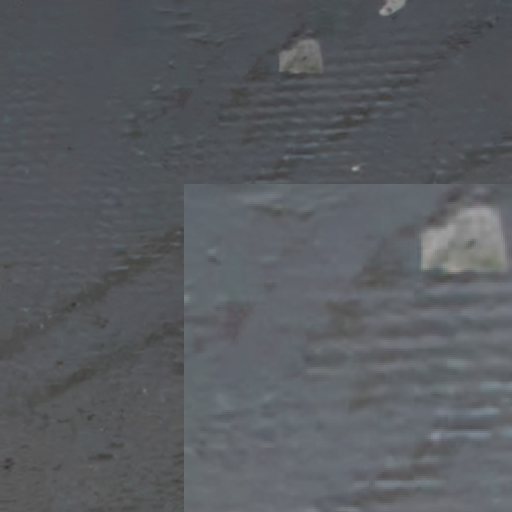}}
\subfigure[CMStSVD]{
\label{Fig4}
\includegraphics[width=0.758in, height = 0.758in]{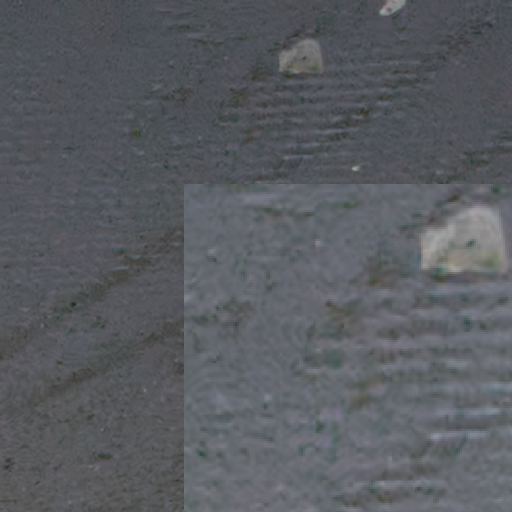}}
\subfigure[CBM3D2]{
\label{Fig4}
\includegraphics[width=0.758in, height = 0.758in]{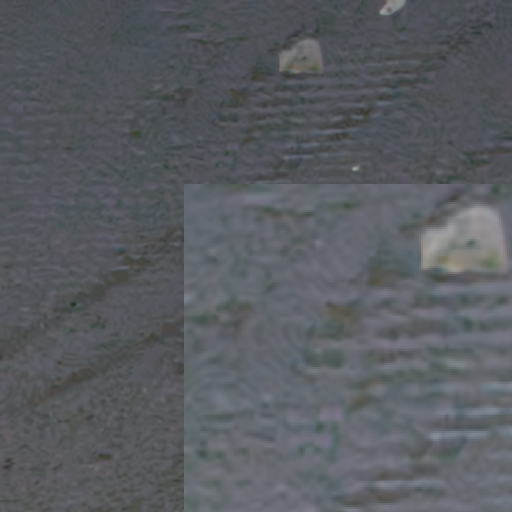}} \\
\vspace{-9.6pt}
\subfigure[NLHCC]{
\label{Fig4}
\includegraphics[width=0.758in, height = 0.758in]{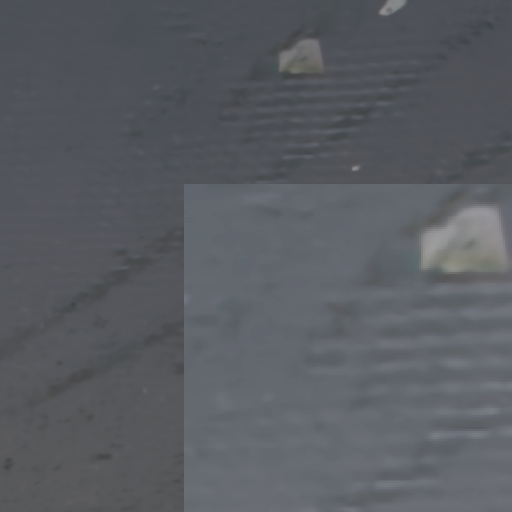}}
\subfigure[FFDNet]{
\label{Fig4}
\includegraphics[width=0.758in, height = 0.758in]{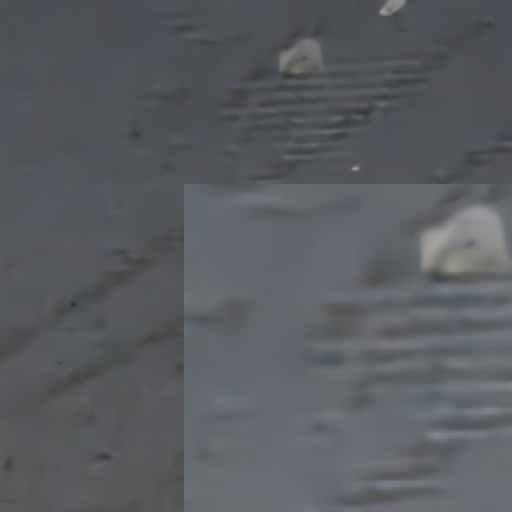}}
\subfigure[SASL]{
\label{Fig4}
\includegraphics[width=0.758in, height = 0.758in]{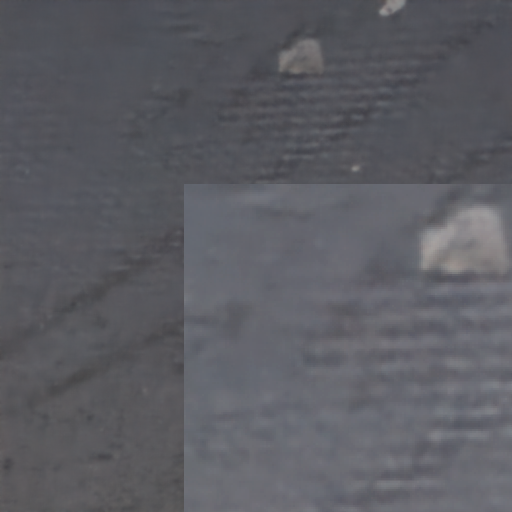}}
\subfigure[Restormer]{
\label{Fig4}
\includegraphics[width=0.758in, height = 0.758in]{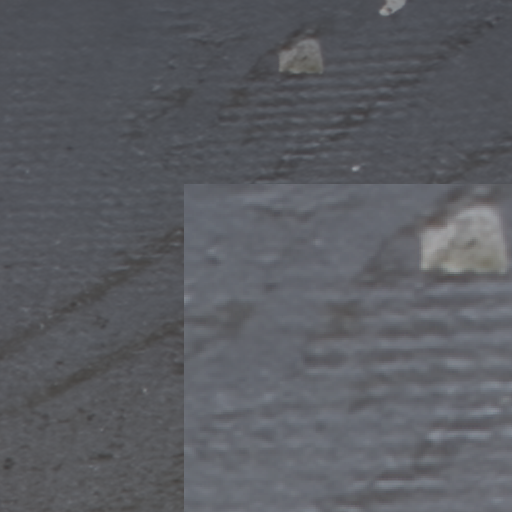}}

\caption{Color Image denoising results on the DND dataset.}
\vspace{-12.8pt}
\label{Fig_DND_visual_evaluation}
\end{figure}

\vspace{-9.9pt}
\indent To better understand and depict the denoising performance of compared methods, visual evaluations are given in Fig. \ref{Fig_CC15_visual_evaluation} and Fig. \ref{Fig_DND_visual_evaluation}. We can observe from Fig. \ref{Fig_CC15_visual_evaluation} that the propose GCP-ID method exhibits competitive capability of both noise removal and detail recovery when the green channel is more important. Interestingly, from Fig. \ref{Fig_CC15_visual_evaluation}, it is noticed that when the green pixels no longer play a dominant role, GCP-ID can still effectively suppress noise and preserve structure information, which demonstrates the benefit of recursively modeling and utilizing the green channel. In addition, in the presence of severe noise, GCP-ID is able to reduce color artifacts and show a comparable performance with the state-of-the-art DNN models.
\vspace{-10pt}


\subsection{Real-world color video denoising}
\vspace{-6.9pt}
\indent \textbf{Datasets.} In this section, two real-world datasets CVRD \cite{yue2020supervised} and IOCV \cite{kong2023comparison} are employed for the color video denoising task. Briefly, CRVD has 55 video clips constructed based on 11 indoor scenes with 5 predefined ISO values. IOCV consists of 39 video clips of various sizes and scenes. The videos are sampled by two different mobile phone devices with auto mode.\\
\indent \textbf{Implementation details.} GCP-ID can be effortlessly extended to handle color videos by performing patch search along both spatial and temporal directions. Similarly, we set $ps = 8$, $W = 16$, $K = 30$, $\tau = 1.1\sigma\sqrt{2\text{log}(3ps^2N_fK)}$, where $N_f$ is the number of frames. The noise level $\sigma$ is chosen from $[25,50]$ with a stepsize of $5$ for both datasets.\\
\indent \textbf{Compared methods.} Similar to color image denoising, both traditional patch-based algorithms and advanced DNN models are included in our evaluations. Since training samples are not provided, we carefully choose decisive parameters and pretrained models of compared methods to obtain their best possible performance during testing.
\begin{table*}[htbp]
\scriptsize
  \centering
   \caption{Average PSNR, SSIM values on two real-world color video datasets. The best results are black bolded.}
   \scalebox{0.868}{
    \begin{tabular}{cccc|cccccccc}
    \toprule
    \multirow{2}[4]{*}{Dataset} & \multicolumn{3}{c|}{Traditional denoisers} & \multicolumn{8}{c}{DNN models} \\
\cmidrule{2-12}           & VBM4D \cite{maggioni2012video} & VIDOSAT \cite{wen2018vidosat} & GCP-ID & DVDNet  \cite{tassano2019dvdnet} & FastDVDNet  \cite{Tassano_2020_CVPR} & FloRNN \cite{li2022unidirectional} & MAP-VDNet \cite{sun2021deep} & RVRT \cite{liang2022recurrent} & UDVD  \cite{sheth2021unsupervised}  & ViDeNN \cite{claus2019videnn} & VNLNet \cite{davy2019non} \\
    \midrule
    \multirow{2}[4]{*}{CRVD} & 34.14  & 34.16  & 36.79  & 34.50  & 35.84  & 36.66  & -     & \textbf{36.94} & -     & 32.31  & 36.11  \\
\cmidrule{2-12}          & 0.9079  & 0.9384  & 0.9511  & 0.9493  & 0.9306  & \textbf{0.9605} & -     & 0.9559  & -     & 0.8449  & 0.9449  \\
    \midrule
    \multirow{2}[4]{*}{IOCV} & 38.76  & -     & \textbf{39.08} & 38.53  & 37.57  & 38.64  & 35.52  & 38.50  & 35.02  & 36.13  & 38.76  \\
\cmidrule{2-12}          & 0.9765  & -     & \textbf{0.9773} & 0.9754  & 0.9699  & 0.9743  & 0.9313  & 0.9672  & 0.9660  & 0.9506  & 0.9765  \\
    \bottomrule
    \end{tabular}}%
    \vspace{-10.8pt}
  \label{Table_color_video_denoising}%
\end{table*}%

\indent \textbf{Performance evaluation.}  Similar to \cite{Tassano_2020_CVPR}, the quality assessment metrics of a sequence are computed as the average of each frame. Table \ref{Table_color_video_denoising} lists the detailed denoising results of compared methods for two different datasets. It can be seen that the proposed GCP-ID is able to produce a comparable denoising performance with powerful DNN models such as FastDVDNet, FloRNN and RVRT on both datasets. Furthermore, compared with representative traditional denoisers, GCP-ID achieves steady improvements in terms of both PSNR and SSIM values. For example, GCP-ID outperforms VBM4D by 2.65 dB and 0.32 dB on CRVD and IOCV, respectively.
\vspace{-10.8pt}
\begin{figure}[htbp]
\graphicspath{{Figs/CRVD/combined/}}
\centering
\subfigure[Reference]{
\label{Fig4}
\includegraphics[width=1.03in, height=1.03in]{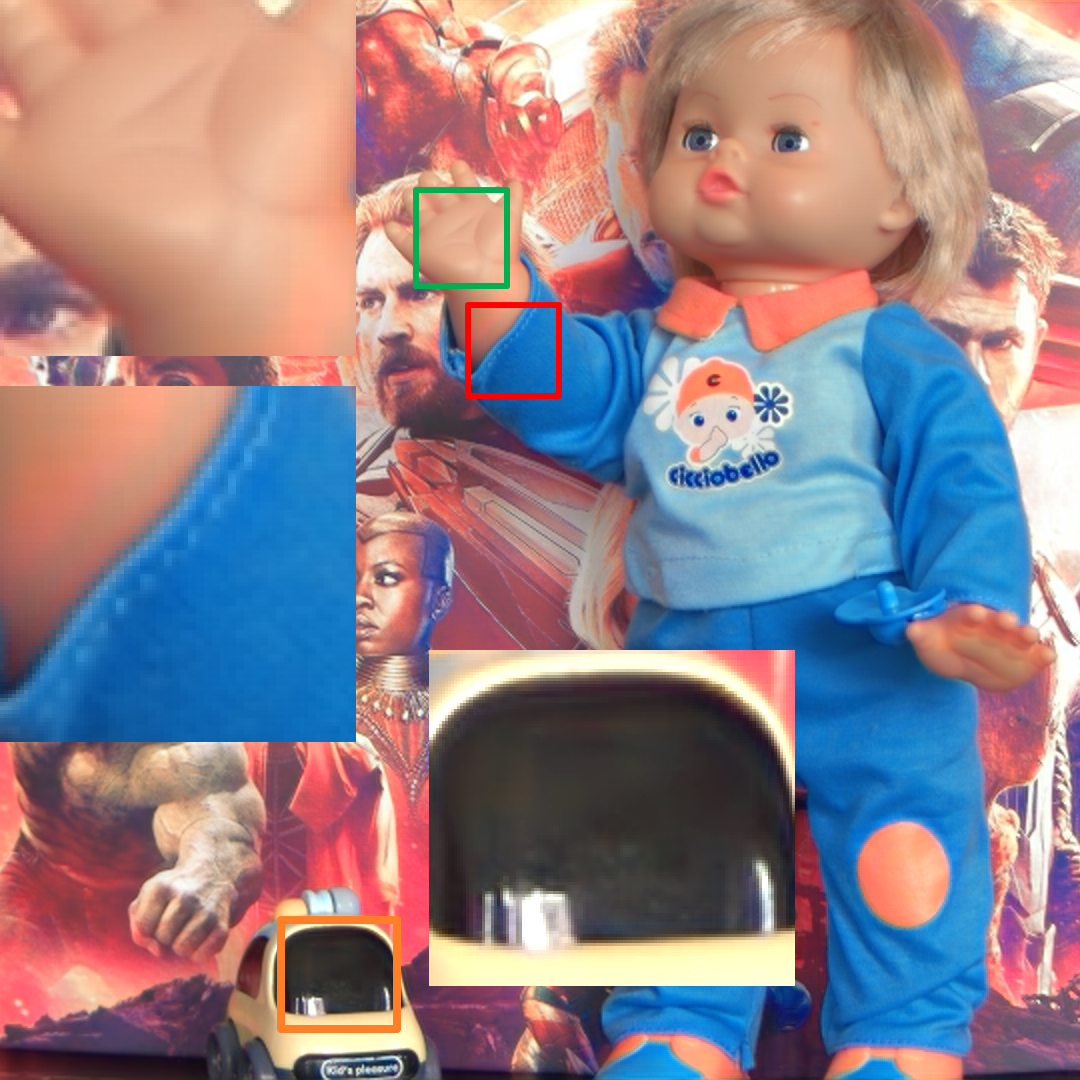}}
\subfigure[Noisy]{
\label{Fig4}
\includegraphics[width=1.03in, height=1.03in]{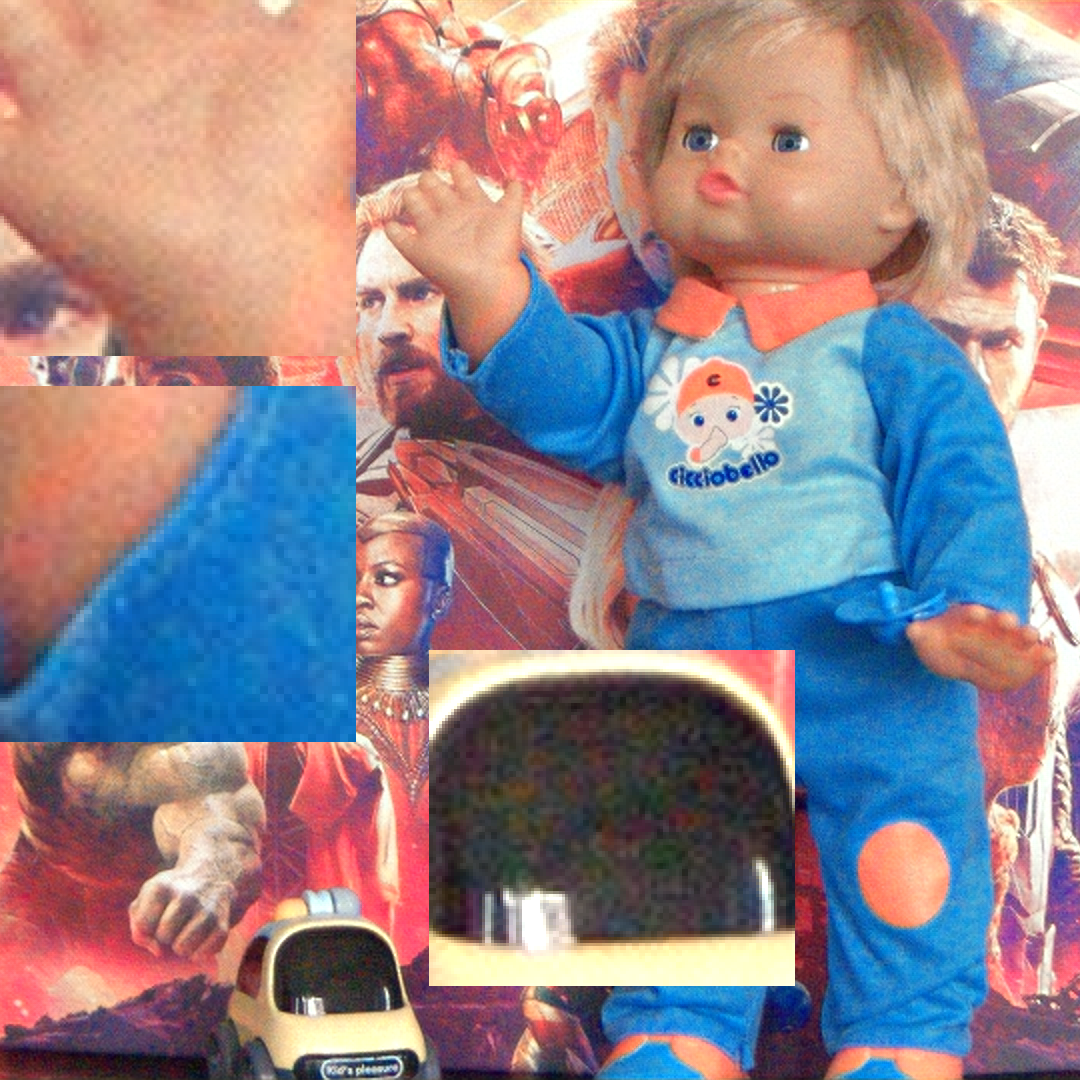}}
\subfigure[GCP-ID]{
\label{Fig4}
\includegraphics[width=1.03in, height=1.03in]{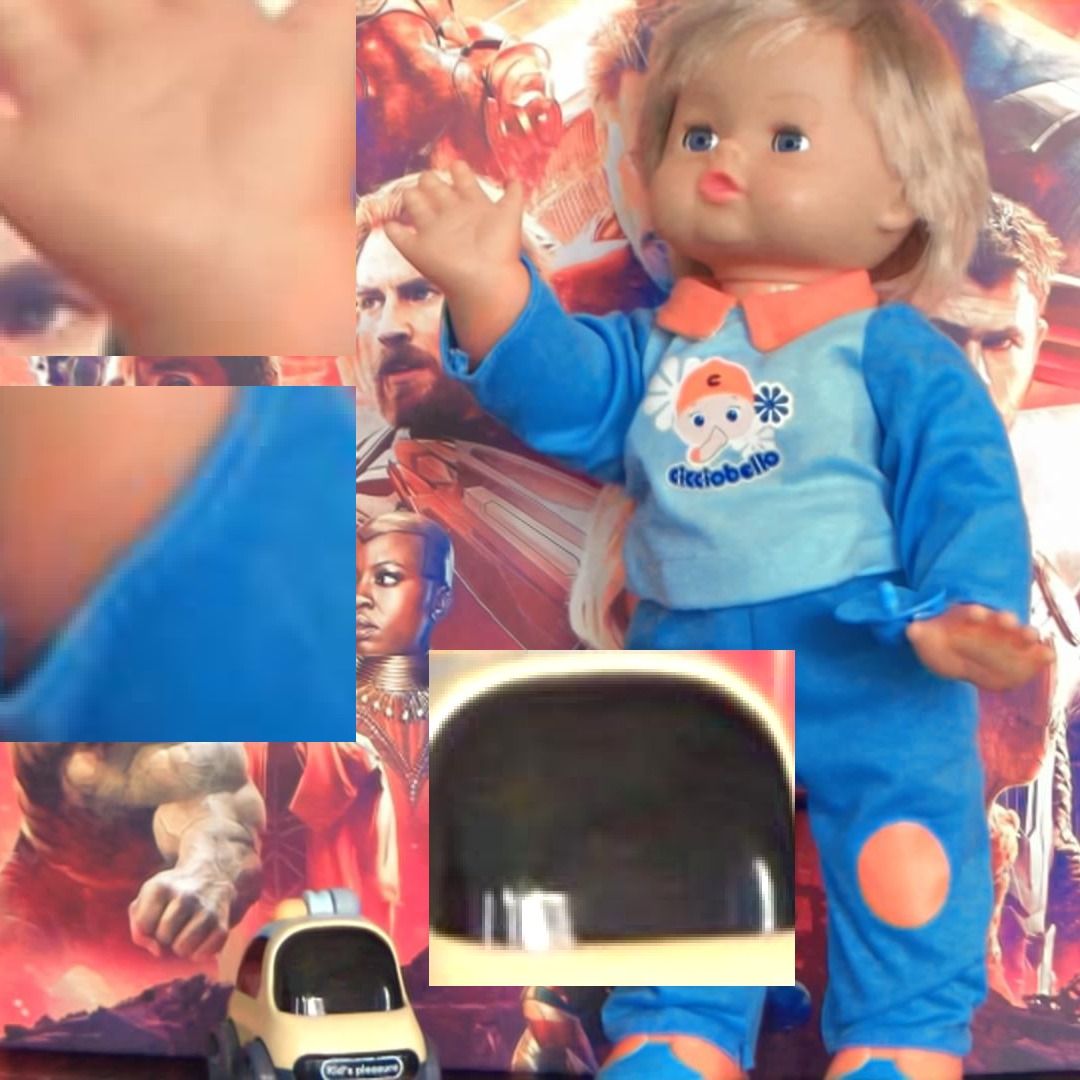}} \\
\vspace{-9.8pt}
\subfigure[VBM4D]{
\label{Fig4}
\includegraphics[width=1.03in, height=1.03in]{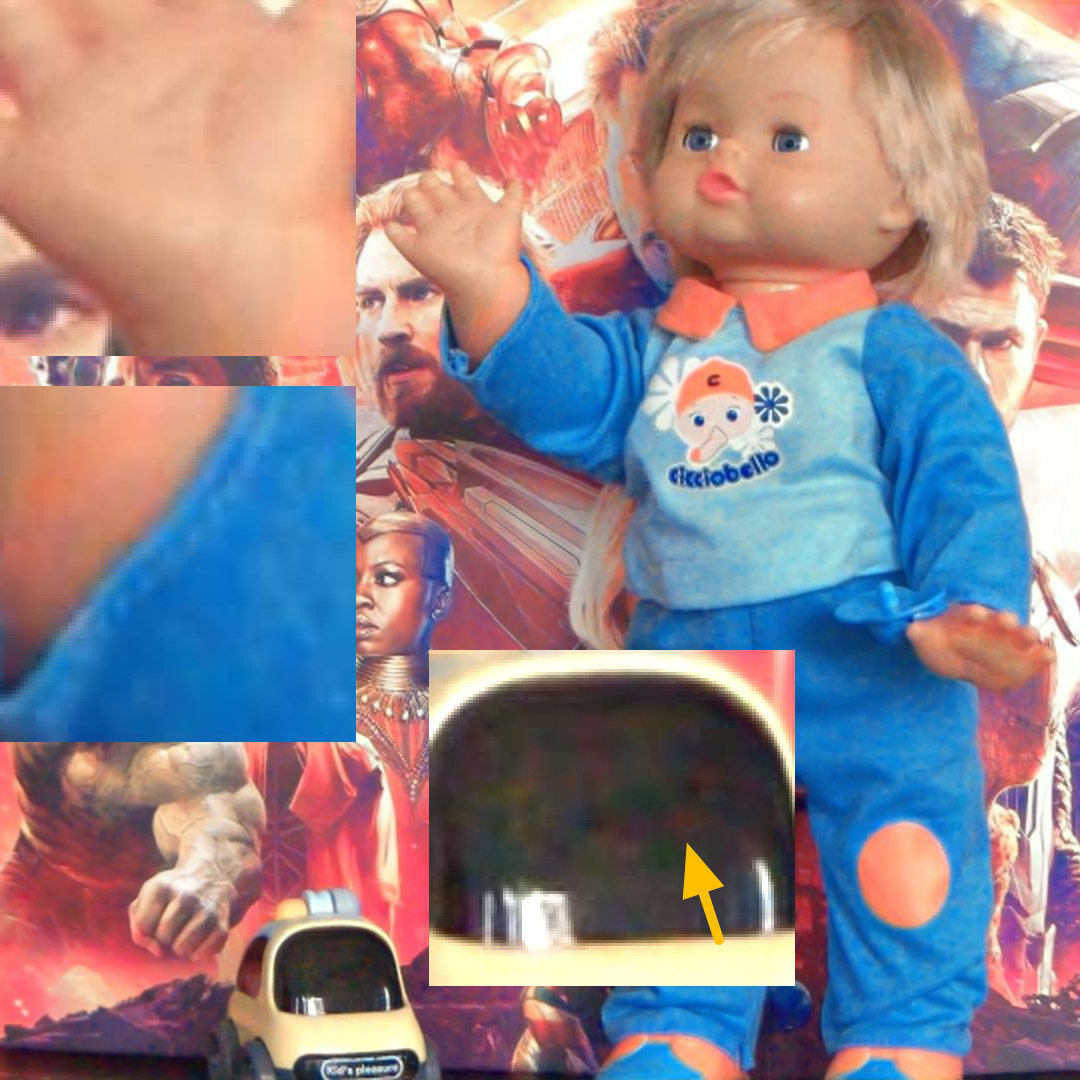}}
\subfigure[FastDVDNet]{
\label{Fig4}
\includegraphics[width=1.03in, height=1.03in]{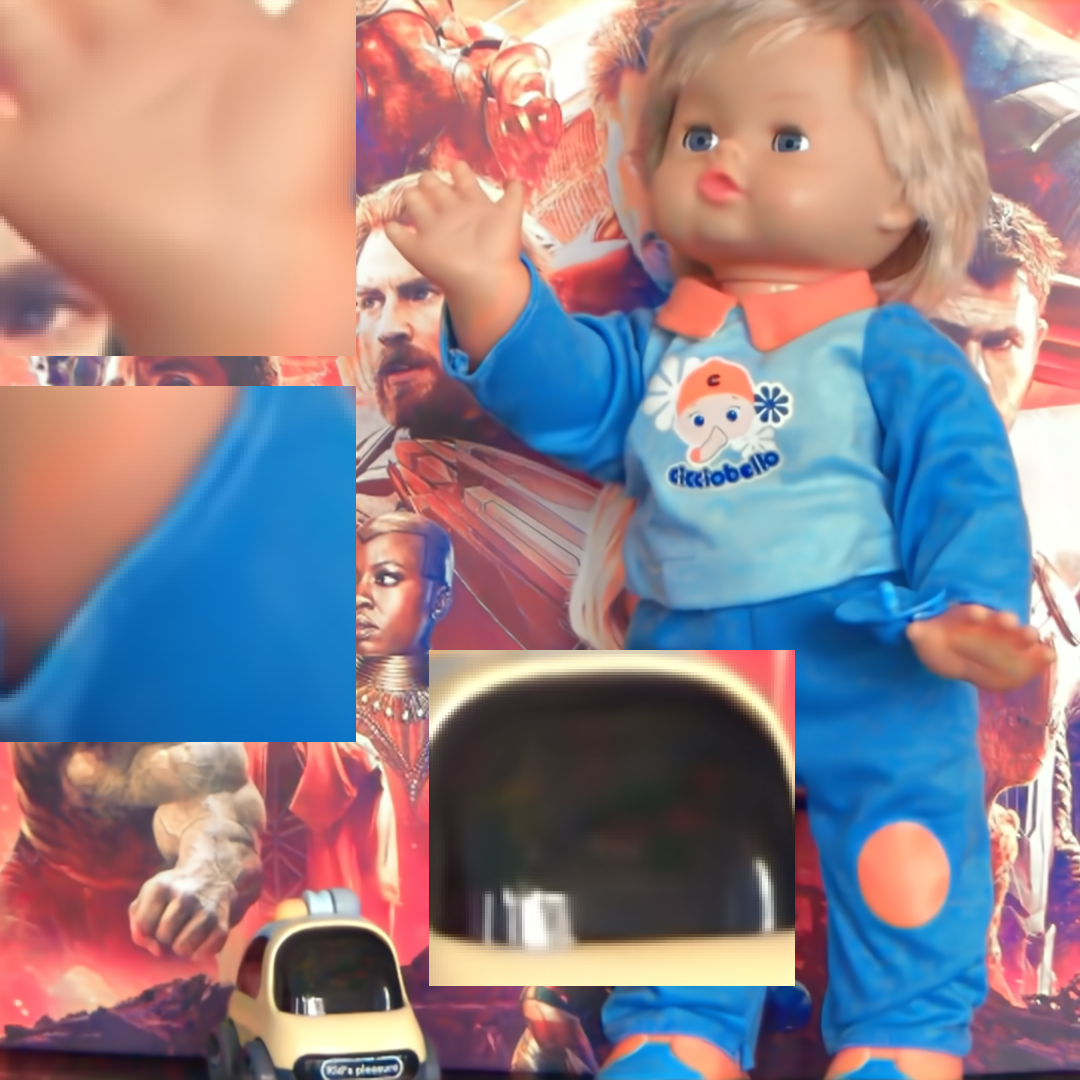}}
\subfigure[RVRT]{
\label{Fig4}
\includegraphics[width=1.03in, height=1.03in]{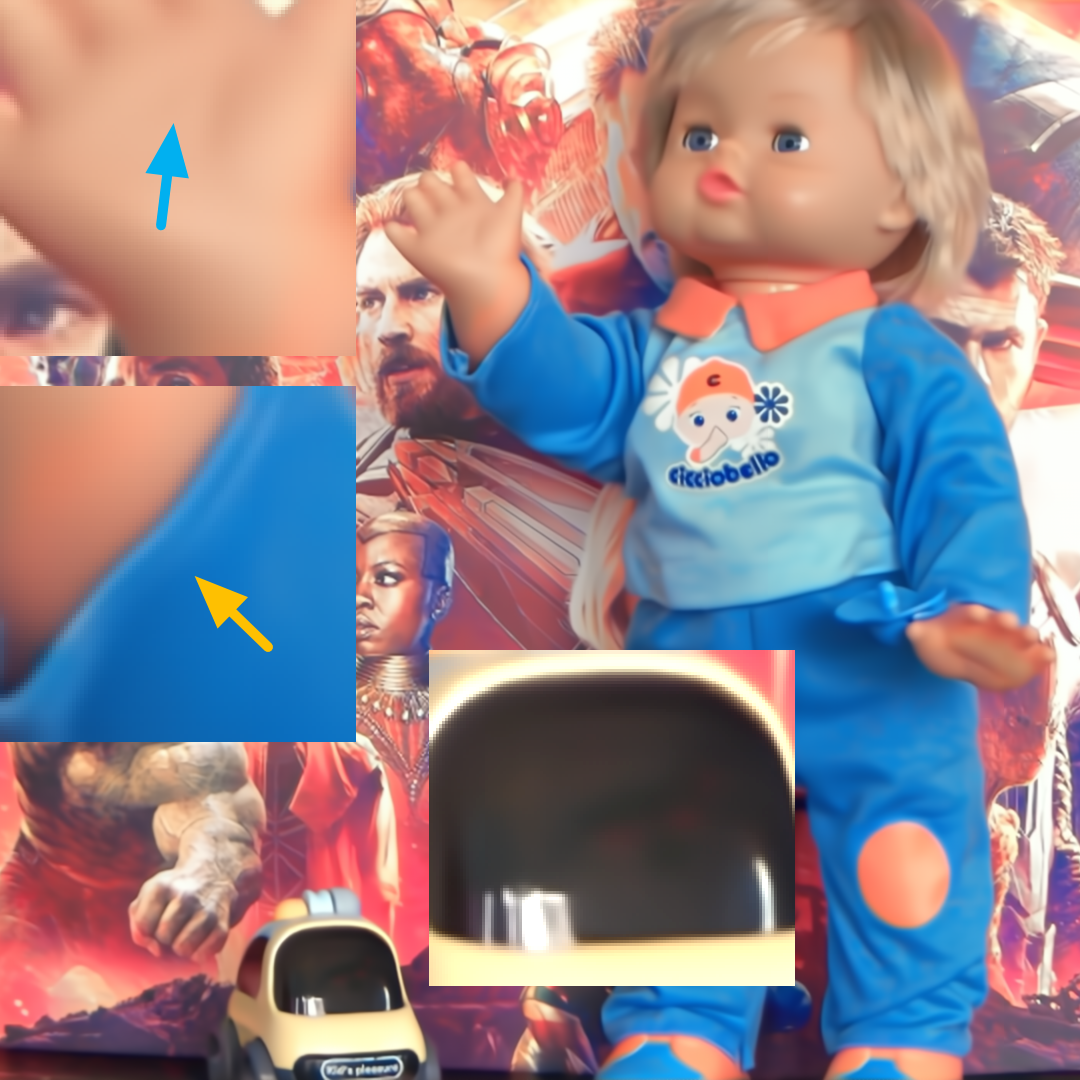}}
\caption{Color video denoising results on the CRVD dataset.}
\label{Fig_CRVD_visual_evaluation}
\vspace{-23.8pt}
\end{figure}

\begin{figure}[htbp]
\graphicspath{{Figs/IOCV/Sample3/combined/}}
\centering
\subfigure[Noisy]{
\label{Fig4}
\includegraphics[width=1.036in, height=1.0in]{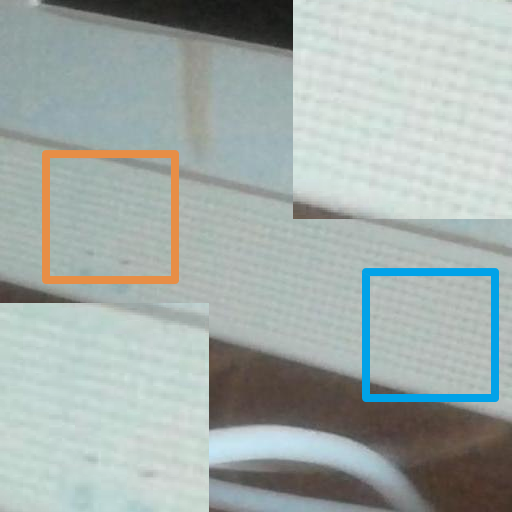}}
\subfigure[GCP-ID]{
\label{Fig4}
\includegraphics[width=1.036in, height=1.0in]{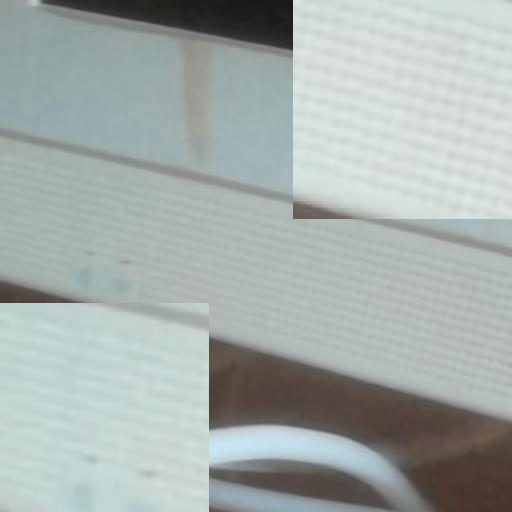}}
\subfigure[VBM4D]{
\label{Fig4}
\includegraphics[width=1.036in, height=1.0in]{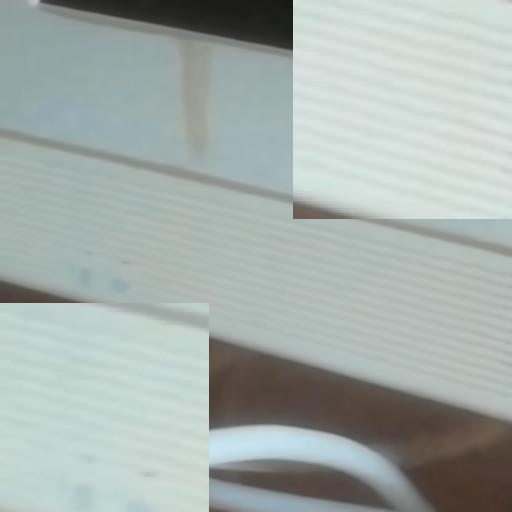}} \\
\vspace{-9.8pt}
\subfigure[FastDVDNet]{
\label{Fig4}
\includegraphics[width=1.036in, height=1.0in]{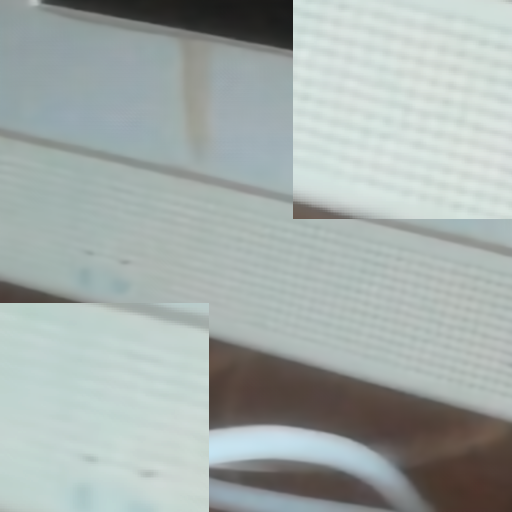}}
\subfigure[FloRNN]{
\label{Fig4}
\includegraphics[width=1.036in, height=1.0in]{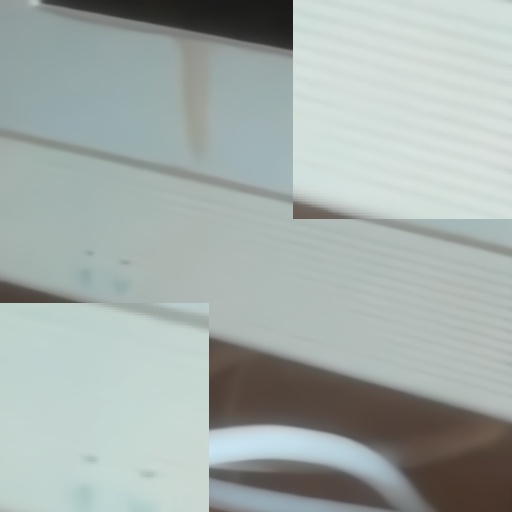}}
\subfigure[RVRT]{
\label{Fig4}
\includegraphics[width=1.036in, height=1.0in]{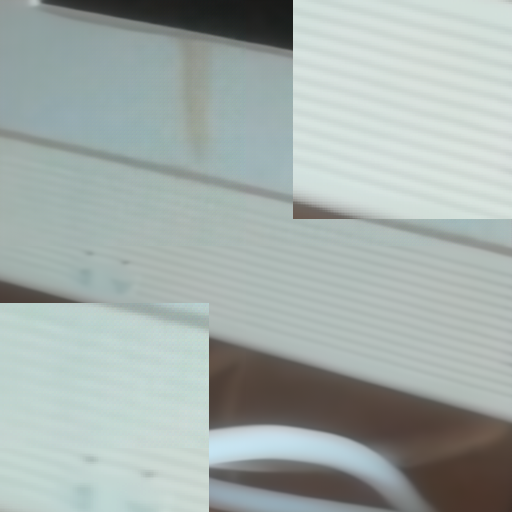}}
\caption{Color video denoising results on the IOCV dataset.}
\label{Fig_IOCV_visual_evaluation}
\vspace{-6.8pt}
\end{figure}

\indent It is worth noting that constructing real-world video datasets is relatively difficult, and thus the reference videos inevitably exhibit noise and motion blur, which will undermine the accuracy of objective evaluations \cite{kong2023comparison}. Therefore, visual quality comparison is provided in Fig. \ref{Fig_CRVD_visual_evaluation} and Fig. \ref{Fig_IOCV_visual_evaluation} to for denoising performance comparison. It can be observed that DNN denoisers FastDVDNet, FloRNN and RVRT demonstrate remarkable temporal coherence thanks to the feature extraction capability of advanced DNN frameworks, traditional patch-based algorithms (\textit{e.g.} GCP-ID and VBM4D) may show notable flickering and leave behind bothersome low-frequency noise in flat parts\cite{Tassano_2020_CVPR}. Meanwhile, we notice that GCP-ID handles textures and structures well, as it benefits from both NLSS prior and repetitive usage of local patch information via RGGB representation.
\subsection{Extension to other imaging techniques}
\begin{figure}[htbp]
\graphicspath{{Figs/Real-HSI/selected/combined/}}
\centering
\subfigure[Reference]{
\label{Fig4}
\includegraphics[width = 1.038in, height = 0.7698in]{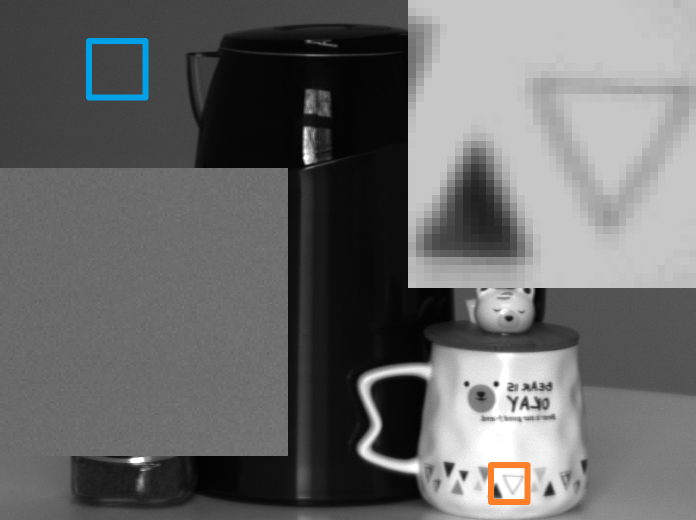}}
\subfigure[Noisy]{
\label{Fig4}
\includegraphics[width = 1.038in, height = 0.7698in]{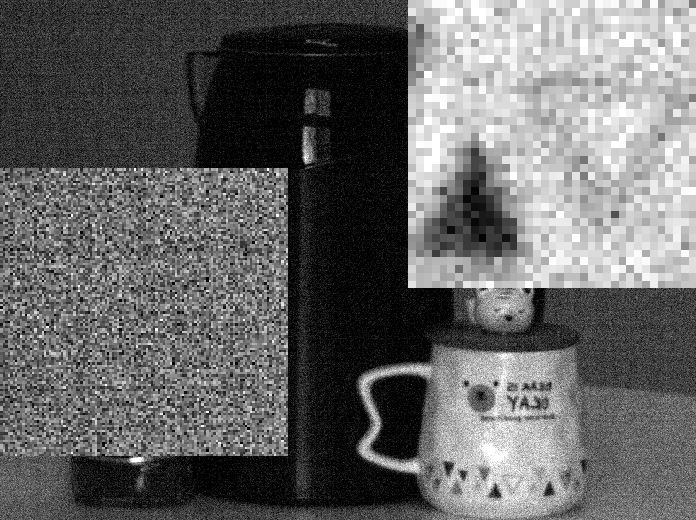}}
\subfigure[GCP-ID]{
\label{Fig4}
\includegraphics[width = 1.038in, height = 0.7698in]{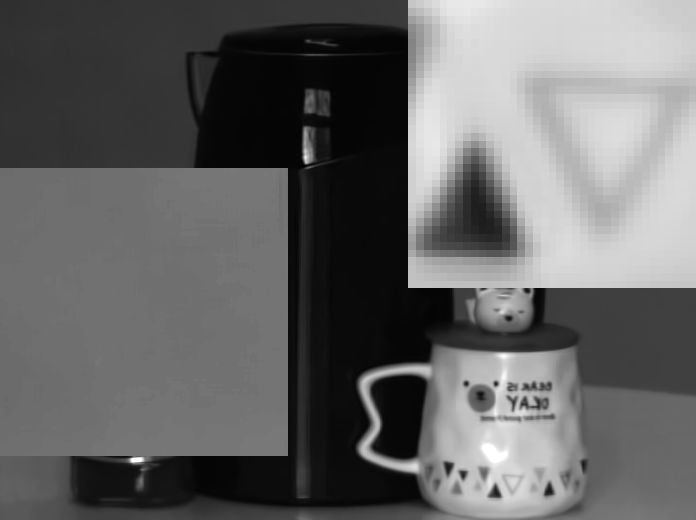}} \\
\vspace{-9.8pt}
\subfigure[OLRT]{
\label{Fig4}
\includegraphics[width = 1.038in, height = 0.7698in]{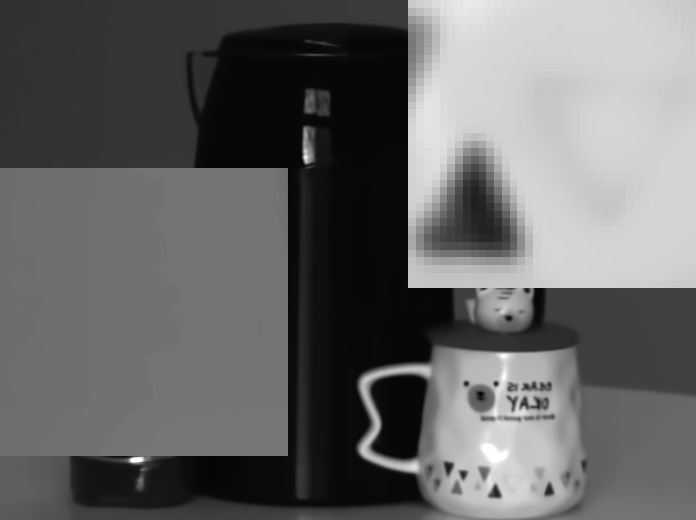}}
\subfigure[MAN]{
\label{Fig4}
\includegraphics[width = 1.038in, height = 0.7698in]{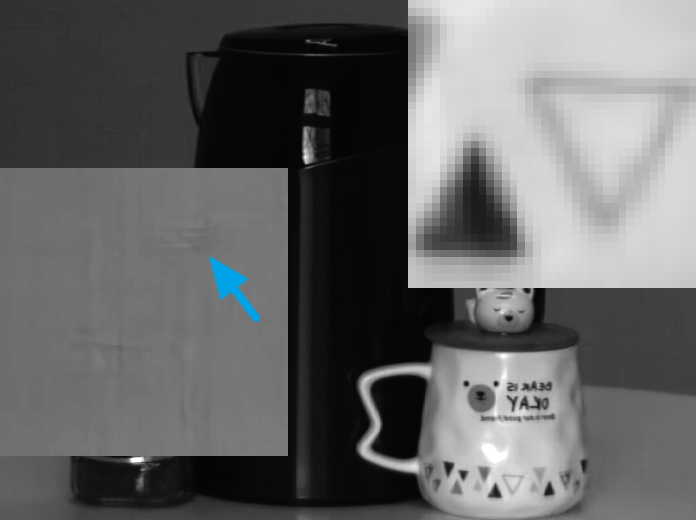}}
\subfigure[QRNN3D]{
\label{Fig4}
\includegraphics[width = 1.038in, height = 0.7698in]{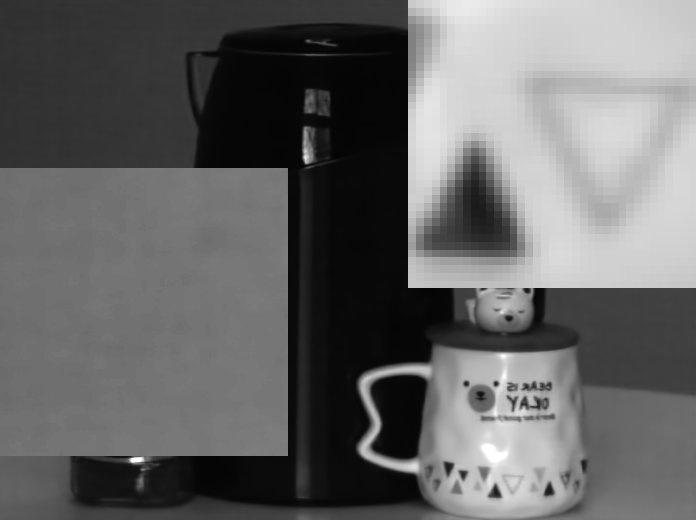}}
\caption{Denoising results on the Real-HSI dataset.}
\label{Fig_Real-HSI_visual_evaluation}
\vspace{-8.8pt}
\end{figure}

The effectiveness of GCP on image and video denoising motivates us to investigate if it may be extended to other imaging techniques, \textit{e.g.}, multispectral and hyperspectral imaging (MSI/HSI). It is noticed that the GCP is based on the sensitivity of human eyes to medium range of spectrum, resulting in the fact that green light at this wavelength produces the impression of highest `brightness' \cite{Gigahertz}. Therefore, such spectrum-wise prior may be borrowed to guide denoising for MSI and HSI. Specifically, given an MSI/HSI input, we first evenly divide the image into $N_g$ groups along the spectral dimension, and spectral bands within the medium spectrum range are used for patch search and nonlocal transform learning according to Eq. (\ref{Equ_summarize_inverse_denoising}). Then all groups of spectral bands share the same similar patch indices and transforms. It should be noted that spectral bands of the MSI/HSI input are treated as equal, since the importance of each spectral band is unknown.\\
\indent To verify the performance of GCP, we use the Real-HSI dataset \cite{zhang2021hyperspectral}, which consists of 59 real clean-noisy HSI pairs of size $696 \times 520 \times 34$. The 34 bands around 400 $nm$ to 700 $nm$ are selected in visible spectral range. For each input, 12 spectral bands (from the 11-th to the 22-th) are selected to search similar patches and learn nonlocal transforms. We set $ps = 8$, $W = 18$, $K = 30$, $N_g = 6$, $\tau = 6\sigma\sqrt{2\text{log}(3ps^2N_sK)}$, where $N_s$ refers to the number of spectral bands for each group. The noise level $\sigma$ is chosen from $[8, 20]$ with a stepsize equal to 2. In addition to PSNR and SSIM, we introduce spectral angle mapper (SAM) \cite{yuhas1990determination} and relative dimensionless global error in synthesis (ERGAS) \cite{wald2002data} for spectral-based quality assessment of HSI data. Recovered data with lower SAM and ERGAS are considered of better quality.
\begin{table*}[htbp]
\scriptsize
  \centering
  \renewcommand{\arraystretch}{0.8198}
  \caption{Comparison of quantitative results on Real-HSI datasets.}
    \begin{tabular}{cccccccccc|cccc}
    \toprule
    \multirow{3}[4]{*}{Datasets} & \multirow{3}[4]{*}{\# Images} & \multirow{3}[4]{*}{Metrics} & \multicolumn{6}{c}{Traditional denoisers}     &       & \multicolumn{4}{c}{DNN models} \\
\cmidrule{4-14}          &       &       & BM4D  & ISTReg & LTDL  & GCP-ID & NGMeet & OLRT  & TDL   & HSI-DeNet & MAN   & QRNN3D & sDeCNN \\
          &       &       & \cite{maggioni2012nonlocal} & \cite{chen2019hyperspectral} & \cite{gong2020low} & (Proposed) & \cite{he2019non} & \cite{chang2020hyperspectral} & \cite{peng2014decomposable} & \cite{chang2018hsi} & \cite{lai2023mixed} & \cite{wei20203} & \cite{maffei2019single} \\
    \midrule
    \multirow{4}[8]{*}{Real-HSI} & \multirow{4}[8]{*}{59} & PSNR  & 25.88  & \textbf{25.91} & 25.80  & \textbf{25.91} & 25.87  & \textbf{25.94} & 25.33  & 25.63  & 25.82  & 25.82  & 25.70  \\
\cmidrule{3-14}          &       & SSIM  & 0.8653  & 0.8688  & 0.8413  & \textbf{0.8695} & 0.8659  & \textbf{0.8695} & 0.7634  & 0.8534  & 0.8694  & 0.8691  & 0.8597  \\
\cmidrule{3-14}          &       & SAM   & 0.0660  & 0.0551  & 0.0742  & 0.0613  & \textbf{0.0509} & 0.0538  & 0.1232  & 0.0921  & 0.0596  & 0.0638  & 0.0928  \\
\cmidrule{3-14}          &       & EGRAS & 222.66  & 222.18  & 223.32  & 222.11  & 222.69  & \textbf{221.94} & 229.32  & 232.73  & 224.41  & 225.32  & 227.88  \\
    \midrule
    Time  & -     & minutes & 4.1   & 41.9  & 35.0  & 3.9   & 4.1   & 24.3  & 0.9   & 0.8   & 0.2   & \textbf{0.1} & 0.7  \\
    \bottomrule
    \end{tabular}%
    \vspace{-15.8pt}
  \label{Table_HSI_results}%
\end{table*}%
\\
\indent The overall quantitative assessment results and average running time by competing denoising methods are reported in Table \ref{Table_HSI_results}. Interestingly, the extension of the proposed GCP-ID is also able to achieve competitive performance for real HSI noise removal, which manifests the importance of leveraging spectrum-wise prior information to guide denoising. From the running time comparison, DNN models shows significant advantage in the testing phase as they benefit from the advanced GPU devices and eliminate the need to learn local transforms for each input. For GCP-ID, since it gets rid of a large number of similar patches and the repetitive calculation of local transforms, its pure MATLAB implementation is more efficient than other state-of-the-art traditional denoisers such as BM4D, LTDL and OLRT. In addition, our C++ MEX functions can further reduce the denoising time by half. \\
\indent Fig. \ref{Fig_Real-HSI_visual_evaluation} visualizes the denoising performance of compared methods. It can be seen that GCP-ID retains the details of small shapes and effectively removes noise in flat areas. The representative low rank method OLRT tends to blur the texture and produces over-smooth effects, because the optimal rank may be hard to obtain. Besides, the pretrained DNN models of MAN and QRNN3D may also leave stripe noise and unwanted artifacts in flat areas.
\vspace{-8pt}

\vspace{-3pt}
\subsection{Discussion}
To handle color images, the proposed GCP-ID mainly consists of two key steps, namely the GCP-guided patch search and the RGGB representation. It is interesting to investigate the difference of each step, thus we use the SIDD validation dataset and compare various implementations of GCP-ID. From Table \ref{Table_discussion_two_step}, we notice that the RGGB representation has a higher PSNR value, while the guided search step is superior in terms of the SSIM metric.
\vspace{-2.8pt}

\begin{table}[htbp]
  \centering
  \caption{The effectiveness of each step of GCP-ID.}
  \scalebox{0.828}{
    \begin{tabular}{ccccc}
    \toprule
    Dataset & Metrics & Guided search & RGGB  & GCP-ID \\
    \midrule
    \multirow{2}[4]{*}{SIDD Validation} & PSNR  & 34.88 & 34.62 & 35.03 \\
\cmidrule{2-5}          & SSIM  & 0.8665 & 0.8738 & 0.8825 \\
    \bottomrule
    \end{tabular}}%
    \vspace{-8pt}
  \label{Table_discussion_two_step}%
\end{table}%

\begin{figure}[htbp]
\graphicspath{{Figs/Discussion/Inpact_two_steps/Selected/img1/}}
\centering
\subfigure[Noisy]{
\label{Fig4}
\includegraphics[width=0.758in, height = 0.758in]{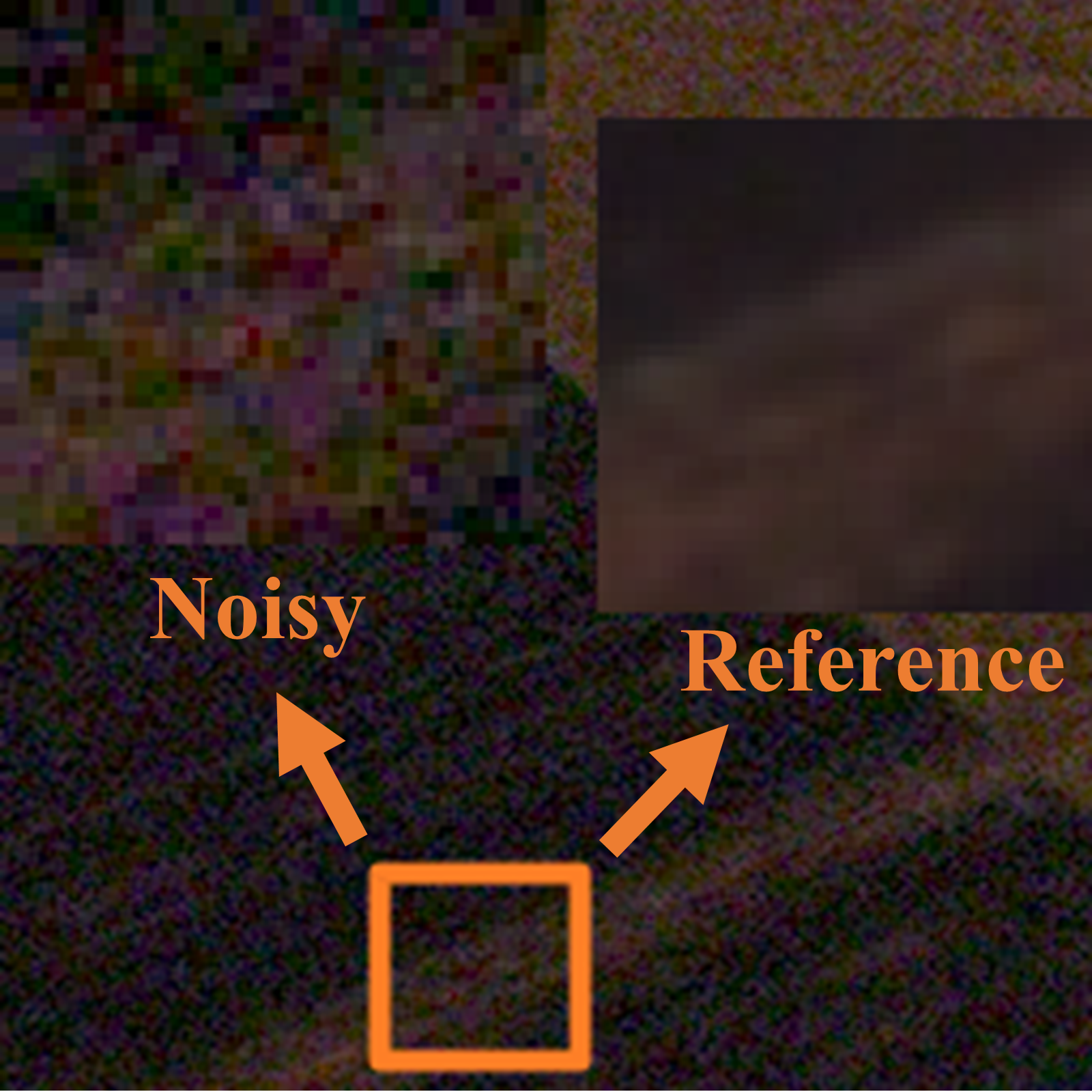}}
\subfigure[Guided]{
\label{Fig4}
\includegraphics[width=0.758in, height = 0.758in]{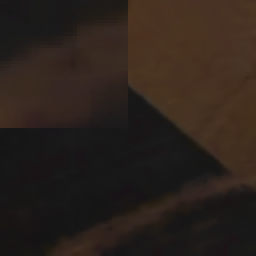}}
\subfigure[RGGB]{
\label{Fig4}
\includegraphics[width=0.758in, height = 0.758in]{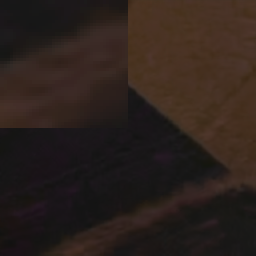}}
\subfigure[GCP-ID]{
\label{Fig4}
\includegraphics[width=0.758in, height = 0.758in]{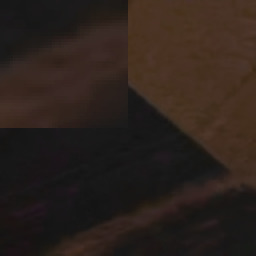}}
\vspace{-6pt}
\graphicspath{{Figs/Discussion/Inpact_two_steps/Selected/img3/}}
\subfigure[Noisy]{
\label{Fig4}
\includegraphics[width=0.75in, height = 0.756in]{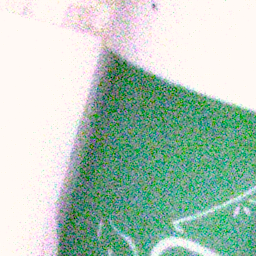}}
\subfigure[Guided]{
\label{Fig4}
\includegraphics[width=0.75in, height = 0.756in]{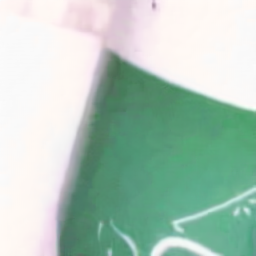}}
\subfigure[RGGB]{
\label{Fig4}
\includegraphics[width=0.75in, height = 0.756in]{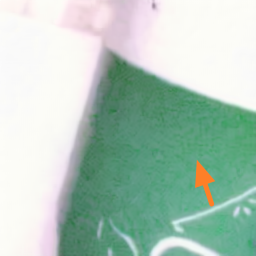}}
\subfigure[GCP-ID]{
\label{Fig4}
\includegraphics[width=0.75in, height = 0.756in]{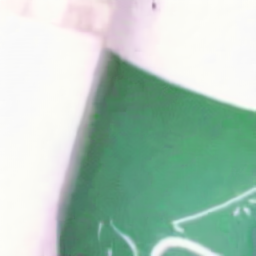}}
\vspace{-1pt}
\caption{Denoising results of different steps of GCP-ID.}
\label{Fig_discussion_guided_search_RGGB}
\end{figure}

\vspace{-0.8pt}
\indent To better understand the observed results, we visualize the denoising performance in Fig. \ref{Fig_discussion_guided_search_RGGB}. It can be seen that the guided search step filters out noise at the cost of more obvious oversmoothness, while the RGGB representation preserves detail and textures but also produces unpleasant artifacts. This interesting observation indicates that the group level redundancy from the GCP-guided similar patch search weighs highly on the sparsity in the transform domain. Furthermore, the patch level redundancy introduced by the recursive use of RGGB retains structural information, meanwhile the noise patterns also repeat, leaving unwanted color artifacts when the image is severely contaminated. As a result, the combination of these two steps by GCP-ID tends to balance noise removal and detail restoration.
\vspace{-11.8pt}

\vspace{-11.8pt}
\section{Conclusion}
In this paper, we present a novel image denoising method termed GCP-ID, which leverages spectrum and channel-wise prior information of image data to guide noise removal. Following the classic denoising paradigm, the proposed GCP-ID integrates the guided search strategy for nonlocal similarity and the RGGB representation for local patches into a unified framework. Experimental results on real-world datasets demonstrate the effectiveness of GCP-ID for different denoising tasks. It is interesting to investigate if the spectrum prior can be leveraged by advanced DNN models for further enhancement.

{\small
\bibliographystyle{ieee_fullname}
\bibliography{reference}

\begin{thebibliography}{10}\itemsep=-1pt

\bibitem{abdelhamed2018high}
Abdelrahman Abdelhamed, Stephen Lin, and Michael~S Brown.
\newblock A high-quality denoising dataset for smartphone cameras.
\newblock In {\em Proc. IEEE Conf. Comput. Vis. Pattern Recognit.}, pages
  1692--1700, 2018.

\bibitem{aharon2006k}
Michal Aharon, Michael Elad, and Alfred Bruckstein.
\newblock K-svd: An algorithm for designing overcomplete dictionaries for
  sparse representation.
\newblock {\em IEEE Trans. Signal Process.}, 54(11):4311--4322, 2006.

\bibitem{anwar2019real}
Saeed Anwar and Nick Barnes.
\newblock Real image denoising with feature attention.
\newblock In {\em Proc. IEEE Int. Conf. Comput. Vis.}, pages 3155--3164, 2019.

\bibitem{beck2009fast}
Amir Beck and Marc Teboulle.
\newblock Fast gradient-based algorithms for constrained total variation image
  denoising and deblurring problems.
\newblock {\em IEEE Trans. Image Process.}, 18(11):2419--2434, 2009.

\bibitem{buades2005review}
Antoni Buades, Bartomeu Coll, and Jean-Michel Morel.
\newblock A review of image denoising algorithms, with a new one.
\newblock {\em Multiscale Model. Simul.}, 4(2):490--530, 2005.

\bibitem{chang2020hyperspectral}
Yi Chang, Luxin Yan, Bingling Chen, Sheng Zhong, and Yonghong Tian.
\newblock Hyperspectral image restoration: Where does the low-rank property
  exist.
\newblock {\em IEEE Trans. Geosci. Remote Sens.}, 59(8):6869--6884, 2020.

\bibitem{chang2018hsi}
Yi Chang, Luxin Yan, Houzhang Fang, Sheng Zhong, and Wenshan Liao.
\newblock Hsi-denet: Hyperspectral image restoration via convolutional neural
  network.
\newblock {\em IEEE Trans. Geosci. Remote Sens.}, 57(2):667--682, 2018.

\bibitem{chang2017hyper}
Yi Chang, Luxin Yan, and Sheng Zhong.
\newblock Hyper-laplacian regularized unidirectional low-rank tensor recovery
  for multispectral image denoising.
\newblock In {\em Proc. IEEE Conf. Comput. Vis. Pattern Recognit.}, pages
  4260--4268, 2017.

\bibitem{chen2019hyperspectral}
Yong Chen, Wei He, Naoto Yokoya, and Ting-Zhu Huang.
\newblock Hyperspectral image restoration using weighted group
  sparsity-regularized low-rank tensor decomposition.
\newblock {\em IEEE Trans. Cybern.}, 50(8):3556--3570, 2019.

\bibitem{chung2008lossless}
King-Hong Chung and Yuk-Hee Chan.
\newblock A lossless compression scheme for bayer color filter array images.
\newblock {\em IEEE Trans. Image Process.}, 17(2):134--144, 2008.

\bibitem{claus2019videnn}
Michele Claus and Jan Van~Gemert.
\newblock Videnn: Deep blind video denoising.
\newblock In {\em Proc. Conf. Comput. Vis. Pattern Recognit. Workshops}, pages
  0--0, 2019.

\bibitem{dabov2007color}
Kostadin Dabov, Alessandro Foi, Vladimir Katkovnik, and Karen Egiazarian.
\newblock Color image denoising via sparse 3d collaborative filtering with
  grouping constraint in luminance-chrominance space.
\newblock In {\em Proc. IEEE Int. Conf. Image Process.}, pages 313--316, 2007.

\bibitem{dabov2007image}
Kostadin Dabov, Alessandro Foi, Vladimir Katkovnik, and Karen Egiazarian.
\newblock Image denoising by sparse 3-d transform-domain collaborative
  filtering.
\newblock {\em IEEE Trans. Image Process.}, 16(8):2080--2095, 2007.

\bibitem{dabov2009bm3d}
Kostadin Dabov, Alessandro Foi, Vladimir Katkovnik, and Karen Egiazarian.
\newblock Bm3d image denoising with shape-adaptive principal component
  analysis.
\newblock In {\em Proc. Workshop on SPARS}, pages 1--6, 2009.

\bibitem{dai2013multichannel}
Jingjing Dai, Oscar~C Au, Lu Fang, Chao Pang, Feng Zou, and Jiali Li.
\newblock Multichannel nonlocal means fusion for color image denoising.
\newblock {\em IEEE Trans. Circuits Syst. Video Technol.}, 23(11):1873--1886,
  2013.

\bibitem{davy2019non}
Axel Davy, Thibaud Ehret, Jean-Michel Morel, Pablo Arias, and Gabriele
  Facciolo.
\newblock A non-local cnn for video denoising.
\newblock In {\em Proc. IEEE Conf. Int. Image Process.}, pages 2409--2413,
  2019.

\bibitem{donoho1994ideal}
David~L Donoho and Jain~M Johnstone.
\newblock Ideal spatial adaptation by wavelet shrinkage.
\newblock {\em biometrika}, 81(3):425--455, 1994.

\bibitem{ehret2017global}
Thibaud Ehret, Pablo Arias, and Jean-Michel Morel.
\newblock Global patch search boosts video denoising.
\newblock In {\em Proc. Int. Conf. Comput. Vis. Theory Appl.}, volume~5, pages
  124--134, 2017.

\bibitem{elad2006image}
Michael Elad and Michal Aharon.
\newblock Image denoising via sparse and redundant representations over learned
  dictionaries.
\newblock {\em IEEE Trans. Image Process.}, 15(12):3736--3745, 2006.

\bibitem{elad2023image}
Michael Elad, Bahjat Kawar, and Gregory Vaksman.
\newblock Image denoising: The deep learning revolution and beyond--a survey
  paper--.
\newblock {\em arXiv preprint arXiv:2301.03362}, 2023.

\bibitem{foi2007pointwise}
Alessandro Foi, Vladimir Katkovnik, and Karen Egiazarian.
\newblock Pointwise shape-adaptive dct for high-quality denoising and
  deblocking of grayscale and color images.
\newblock {\em IEEE Trans. Image Process.}, 16(5):1395--1411, 2007.

\bibitem{Gigahertz}
Gigahertz-Optik.
\newblock Spectral sensitivity of the human eye.
\newblock \url{https://light-measurement.com/spectral-sensitivity-of-eye}.

\bibitem{gong2020low}
Xiao Gong, Wei Chen, and Jie Chen.
\newblock A low-rank tensor dictionary learning method for hyperspectral image
  denoising.
\newblock {\em IEEE Trans. Signal Process.}, 68:1168--1180, 2020.

\bibitem{guo2021joint}
Shi Guo, Zhetong Liang, and Lei Zhang.
\newblock Joint denoising and demosaicking with green channel prior for
  real-world burst images.
\newblock {\em IEEE Trans. Image Process.}, 30:6930--6942, 2021.

\bibitem{guo2019toward}
Shi Guo, Zifei Yan, Kai Zhang, Wangmeng Zuo, and Lei Zhang.
\newblock Toward convolutional blind denoising of real photographs.
\newblock In {\em Proc. Conf. Comput. Vis. Pattern Recognit.}, pages
  1712--1722, 2019.

\bibitem{he2019non}
Wei He, Quanming Yao, Chao Li, Naoto Yokoya, and Qibin Zhao.
\newblock Non-local meets global: An integrated paradigm for hyperspectral
  denoising.
\newblock In {\em Proc. IEEE Conf. Comput. Vis. Pattern Recognit.}, pages
  6861--6870, 2019.

\bibitem{hou2020nlh}
Yingkun Hou, Jun Xu, Mingxia Liu, Guanghai Liu, Li Liu, Fan Zhu, and Ling Shao.
\newblock Nlh: A blind pixel-level non-local method for real-world image
  denoising.
\newblock {\em IEEE Trans. Image Process.}, 29:5121--5135, 2020.

\bibitem{huang2021neighbor2neighbor}
Tao Huang, Songjiang Li, Xu Jia, Huchuan Lu, and Jianzhuang Liu.
\newblock Neighbor2neighbor: Self-supervised denoising from single noisy
  images.
\newblock In {\em Proc. IEEE Conf. Comput. Vis. Pattern Recognit.}, pages
  14781--14790, 2021.

\bibitem{jang2021c2n}
Geonwoon Jang, Wooseok Lee, Sanghyun Son, and Kyoung~Mu Lee.
\newblock C2n: Practical generative noise modeling for real-world denoising.
\newblock In {\em Proc. IEEE Int. Conf. Comput. Vis.}, pages 2350--2359, 2021.

\bibitem{jha2010denoising}
Sunil~K Jha and RDS Yadava.
\newblock Denoising by singular value decomposition and its application to
  electronic nose data processing.
\newblock {\em IEEE Sens. J.}, 11(1):35--44, 2010.

\bibitem{kilmer2013third}
Misha~E Kilmer, Karen Braman, Ning Hao, and Randy~C Hoover.
\newblock Third-order tensors as operators on matrices: A theoretical and
  computational framework with applications in imaging.
\newblock {\em SIAM J. Matrix Anal. Appl.}, 34(1):148--172, 2013.

\bibitem{kilmer2011factorization}
Misha~E Kilmer and Carla~D Martin.
\newblock Factorization strategies for third-order tensors.
\newblock {\em Linear Algebra Appl.}, 435(3):641--658, 2011.

\bibitem{Kim_2020_CVPR}
Yoonsik Kim, Jae~Woong Soh, Gu~Yong Park, and Nam~Ik Cho.
\newblock Transfer learning from synthetic to real-noise denoising with
  adaptive instance normalization.
\newblock In {\em Proc. IEEE Conf. Comput. Vis. Pattern Recognit.}, pages
  3482--3492, 2020.

\bibitem{kolda2009tensor}
Tamara~G Kolda and Brett~W Bader.
\newblock Tensor decompositions and applications.
\newblock {\em SIAM Rev.}, 51(3):455--500, 2009.

\bibitem{kong2023comparison}
Zhaoming Kong, Fangxi Deng, Haomin Zhuang, Xiaowei Yang, Jun Yu, and Lifang He.
\newblock A comparison of image denoising methods.
\newblock {\em arXiv preprint arXiv:2304.08990}, 2023.

\bibitem{kong2019color}
Zhaoming Kong and Xiaowei Yang.
\newblock Color image and multispectral image denoising using block diagonal
  representation.
\newblock {\em IEEE Trans. Image Process.}, 28(9):4247--4259, 2019.

\bibitem{lai2023mixed}
Zeqiang Lai and Ying Fu.
\newblock Mixed attention network for hyperspectral image denoising.
\newblock {\em arXiv preprint arXiv:2301.11525}, 2023.

\bibitem{lee2022ap}
Wooseok Lee, Sanghyun Son, and Kyoung~Mu Lee.
\newblock Ap-bsn: Self-supervised denoising for real-world images via
  asymmetric pd and blind-spot network.
\newblock In {\em Proc. IEEE Conf. Comput. Vis. Pattern Recognit.}, pages
  17725--17734, 2022.

\bibitem{li2022unidirectional}
Junyi Li, Xiaohe Wu, Zhenxing Niu, and Wangmeng Zuo.
\newblock Unidirectional video denoising by mimicking backward recurrent
  modules with look-ahead forward ones.
\newblock In {\em Proc. Eur. Conf. Comput. Vis.}, pages 592--609. Springer,
  2022.

\bibitem{li2023spatially}
Junyi Li, Zhilu Zhang, Xiaoyu Liu, Chaoyu Feng, Xiaotao Wang, Lei Lei, and
  Wangmeng Zuo.
\newblock Spatially adaptive self-supervised learning for real-world image
  denoising.
\newblock In {\em Proc. IEEE Conf. Comput. Vis. Pattern Recognit.}, 2023.

\bibitem{li2008image}
Xin Li, Bahadir Gunturk, and Lei Zhang.
\newblock Image demosaicing: A systematic survey.
\newblock In {\em Vis. commun. image process.}, volume 6822, pages 489--503.
  SPIE, 2008.

\bibitem{liang2021swinir}
Jingyun Liang, Jiezhang Cao, Guolei Sun, Kai Zhang, Luc Van~Gool, and Radu
  Timofte.
\newblock Swinir: Image restoration using swin transformer.
\newblock In {\em Proc. IEEE Int. Conf. Comput. Vis.}, pages 1833--1844, 2021.

\bibitem{liang2022recurrent}
Jingyun Liang, Yuchen Fan, Xiaoyu Xiang, Rakesh Ranjan, Eddy Ilg, Simon Green,
  Jiezhang Cao, Kai Zhang, Radu Timofte, and Luc~V Gool.
\newblock Recurrent video restoration transformer with guided deformable
  attention.
\newblock {\em Proc. Advances Neural Inf. Process. Syst.}, 35:378--393, 2022.

\bibitem{liu2020joint}
Lin Liu, Xu Jia, Jianzhuang Liu, and Qi Tian.
\newblock Joint demosaicing and denoising with self guidance.
\newblock In {\em Proc. IEEE Conf. Comput. Vis. Pattern Recognit.}, pages
  2240--2249, 2020.

\bibitem{liu2021invertible}
Yang Liu, Zhenyue Qin, Saeed Anwar, Pan Ji, Dongwoo Kim, Sabrina Caldwell, and
  Tom Gedeon.
\newblock Invertible denoising network: A light solution for real noise
  removal.
\newblock In {\em Proc. IEEE Conf. Comput. Vis. Pattern Recognit.}, pages
  13365--13374, 2021.

\bibitem{maffei2019single}
Alessandro Maffei, Juan~M Haut, Mercedes~Eugenia Paoletti, Javier Plaza,
  Lorenzo Bruzzone, and Antonio Plaza.
\newblock A single model cnn for hyperspectral image denoising.
\newblock {\em IEEE Trans. Geosci. Remote Sens.}, 58(4):2516--2529, 2019.

\bibitem{maggioni2012video}
Matteo Maggioni, Giacomo Boracchi, Alessandro Foi, and Karen Egiazarian.
\newblock Video denoising, deblocking, and enhancement through separable 4-d
  nonlocal spatiotemporal transforms.
\newblock {\em IEEE Trans. Image Process.}, 21(9):3952--3966, 2012.

\bibitem{maggioni2012nonlocal}
Matteo Maggioni, Vladimir Katkovnik, Karen Egiazarian, and Alessandro Foi.
\newblock Nonlocal transform-domain filter for volumetric data denoising and
  reconstruction.
\newblock {\em IEEE Trans. Image Process.}, 22(1):119--133, 2012.

\bibitem{mahdaoui2022image}
Assia~El Mahdaoui, Abdeldjalil Ouahabi, and Mohamed~Said Moulay.
\newblock Image denoising using a compressive sensing approach based on
  regularization constraints.
\newblock {\em Sensors}, 22(6):2199, 2022.

\bibitem{menon2011color}
Daniele Menon and Giancarlo Calvagno.
\newblock Color image demosaicking: An overview.
\newblock {\em Signal Process. Image Commun.}, 26(8-9):518--533, 2011.

\bibitem{nam2016holistic}
Seonghyeon Nam, Youngbae Hwang, Yasuyuki Matsushita, and Seon Joo~Kim.
\newblock A holistic approach to cross-channel image noise modeling and its
  application to image denoising.
\newblock In {\em Proc. IEEE Conf. Comput. Vis. Pattern Recognit.}, pages
  1683--1691, 2016.

\bibitem{pang2017graph}
Jiahao Pang and Gene Cheung.
\newblock Graph laplacian regularization for image denoising: Analysis in the
  continuous domain.
\newblock {\em IEEE Trans. Image Process.}, 26(4):1770--1785, 2017.

\bibitem{peng2014decomposable}
Yi Peng, Deyu Meng, Zongben Xu, Chenqiang Gao, Yi Yang, and Biao Zhang.
\newblock Decomposable nonlocal tensor dictionary learning for multispectral
  image denoising.
\newblock In {\em Proc. IEEE Conf. Comput. Vis. Pattern Recognit.}, pages
  2949--2956, 2014.

\bibitem{plotz2017benchmarking}
Tobias Plotz and Stefan Roth.
\newblock Benchmarking denoising algorithms with real photographs.
\newblock In {\em Proc. IEEE Conf. Comput. Vis. Pattern Recognit.}, pages
  1586--1595, 2017.

\bibitem{quan2020self2self}
Yuhui Quan, Mingqin Chen, Tongyao Pang, and Hui Ji.
\newblock Self2self with dropout: Learning self-supervised denoising from
  single image.
\newblock In {\em Proc. IEEE Conf. Comput. Vis. Pattern Recognit.}, pages
  1890--1898, 2020.

\bibitem{rajwade2012image}
Ajit Rajwade, Anand Rangarajan, and Arunava Banerjee.
\newblock Image denoising using the higher order singular value decomposition.
\newblock {\em IEEE Trans. Pattern Anal. Mach. Intell.}, 35(4):849--862, 2012.

\bibitem{ren2021adaptive}
Chao Ren, Xiaohai He, Chuncheng Wang, and Zhibo Zhao.
\newblock Adaptive consistency prior based deep network for image denoising.
\newblock In {\em Proc. IEEE Conf. Comput. Vis. Pattern Recognit.}, pages
  8596--8606, 2021.

\bibitem{ruikar2011wavelet}
Sachin~D Ruikar and Dharmpal~D Doye.
\newblock Wavelet based image denoising technique.
\newblock {\em Int. J. Adv. Comput. Sci. Appl.}, 2(3), 2011.

\bibitem{sheth2021unsupervised}
Dev~Yashpal Sheth, Sreyas Mohan, Joshua~L Vincent, Ramon Manzorro, Peter~A
  Crozier, Mitesh~M Khapra, Eero~P Simoncelli, and Carlos Fernandez-Granda.
\newblock Unsupervised deep video denoising.
\newblock In {\em Proc. IEEE Int. Conf. Comput. Vis.}, pages 1759--1768, 2021.

\bibitem{shi2021robust}
Qiquan Shi, Yiu-Ming Cheung, and Jian Lou.
\newblock Robust tensor svd and recovery with rank estimation.
\newblock {\em IEEE Trans. Cybern.}, 52(10):10667--10682, 2021.

\bibitem{soh2022variational}
Jae~Woong Soh and Nam~Ik Cho.
\newblock Variational deep image restoration.
\newblock {\em IEEE Trans. Image Process.}, 31:4363--4376, 2022.

\bibitem{sun2021deep}
Lu Sun, Weisheng Dong, Xin Li, Jinjian Wu, Leida Li, and Guangming Shi.
\newblock Deep maximum a posterior estimator for video denoising.
\newblock {\em Int. J. Comput. Vis.}, 129:2827--2845, 2021.

\bibitem{tassano2019dvdnet}
Matias Tassano, Julie Delon, and Thomas Veit.
\newblock Dvdnet: A fast network for deep video denoising.
\newblock In {\em Proc. IEEE Conf. Int. Image Process.}, pages 1805--1809,
  2019.

\bibitem{Tassano_2020_CVPR}
Matias Tassano, Julie Delon, and Thomas Veit.
\newblock Fastdvdnet: Towards real-time deep video denoising without flow
  estimation.
\newblock In {\em Proc. IEEE Conf. Comput. Vis. Pattern Recognit.}, June 2020.

\bibitem{tian2023multi}
Chunwei Tian, Menghua Zheng, Wangmeng Zuo, Bob Zhang, Yanning Zhang, and David
  Zhang.
\newblock Multi-stage image denoising with the wavelet transform.
\newblock {\em Pattern Recognit.}, 134:109050, 2023.

\bibitem{treece2022real}
Graham Treece.
\newblock Real image denoising with a locally-adaptive bitonic filter.
\newblock {\em IEEE Trans. Image Process.}, 31:3151--3165, 2022.

\bibitem{wald2002data}
Lucien Wald.
\newblock {\em Data fusion: definitions and architectures: fusion of images of
  different spatial resolutions}.
\newblock Les Presses de l'Ecoledes Mines, 2002.

\bibitem{wang2004image}
Zhou Wang, Alan~C Bovik, Hamid~R Sheikh, and Eero~P Simoncelli.
\newblock Image quality assessment: from error visibility to structural
  similarity.
\newblock {\em IEEE Trans. Image Process.}, 13(4):600--612, 2004.

\bibitem{wei20203}
Kaixuan Wei, Ying Fu, and Hua Huang.
\newblock 3-d quasi-recurrent neural network for hyperspectral image denoising.
\newblock {\em IEEE Trans. Neural Netw. Learn. Syst.}, 2020.

\bibitem{wen2018vidosat}
Bihan Wen, Saiprasad Ravishankar, and Yoram Bresler.
\newblock Vidosat: High-dimensional sparsifying transform learning for online
  video denoising.
\newblock {\em IEEE Trans. Image Process.}, 28(4):1691--1704, 2018.

\bibitem{xu2018real}
Jun Xu, Hui Li, Zhetong Liang, David Zhang, and Lei Zhang.
\newblock Real-world noisy image denoising: A new benchmark.
\newblock {\em arXiv preprint arXiv:1804.02603}, 2018.

\bibitem{xu2018trilateral}
Jun Xu, Lei Zhang, and David Zhang.
\newblock A trilateral weighted sparse coding scheme for real-world image
  denoising.
\newblock In {\em Proc. Eur. Conf. Comput. Vis.}, pages 20--36, 2018.

\bibitem{xu2017multi}
Jun Xu, Lei Zhang, David Zhang, and Xiangchu Feng.
\newblock Multi-channel weighted nuclear norm minimization for real color image
  denoising.
\newblock In {\em Proc. IEEE Int. Conf. Comput. Vis.}, pages 1096--1104, 2017.

\bibitem{xu2015patch}
Jun Xu, Lei Zhang, Wangmeng Zuo, David Zhang, and Xiangchu Feng.
\newblock Patch group based nonlocal self-similarity prior learning for image
  denoising.
\newblock In {\em Proc. IEEE Conf. Comput. Vis.}, pages 244--252, 2015.

\bibitem{yaroslavsky2001transform}
Leonid~P Yaroslavsky, Karen~O Egiazarian, and Jaakko~T Astola.
\newblock Transform domain image restoration methods: review, comparison, and
  interpretation.
\newblock In {\em Proc. Nonlinear Image Process. Pattern Anal. XII}, volume
  4304, pages 155--169, 2001.

\bibitem{yu2019deep}
Songhyun Yu, Bumjun Park, and Jechang Jeong.
\newblock Deep iterative down-up cnn for image denoising.
\newblock In {\em Proc. Conf. Comput. Vis. Pattern Recognit. Workshops}, 2019.

\bibitem{yue2020supervised}
Huanjing Yue, Cong Cao, Lei Liao, Ronghe Chu, and Jingyu Yang.
\newblock Supervised raw video denoising with a benchmark dataset on dynamic
  scenes.
\newblock In {\em Proc. IEEE Conf. Comput. Vis. Pattern Recognit.}, pages
  2301--2310, 2020.

\bibitem{yue2019high}
Huanjing Yue, Jianjun Liu, Jingyu Yang, Truong~Q Nguyen, and Feng Wu.
\newblock High iso jpeg image denoising by deep fusion of collaborative and
  convolutional filtering.
\newblock {\em IEEE Trans. Image Process.}, 28(9):4339--4353, 2019.

\bibitem{yuhas1990determination}
Roberta~H Yuhas, Joseph~W Boardman, and Alexander~FH Goetz.
\newblock Determination of semi-arid landscape endmembers and seasonal trends
  using convex geometry spectral unmixing techniques.
\newblock {\em ratio}, 4:22, 1990.

\bibitem{zamir2022restormer}
Syed~Waqas Zamir, Aditya Arora, Salman Khan, Munawar Hayat, Fahad~Shahbaz Khan,
  and Ming-Hsuan Yang.
\newblock Restormer: Efficient transformer for high-resolution image
  restoration.
\newblock In {\em Proc. IEEE Conf. Comput. Vis. Pattern Recognit.}, pages
  5728--5739, 2022.

\bibitem{zamir2020cycleisp}
Syed~Waqas Zamir, Aditya Arora, Salman Khan, Munawar Hayat, Fahad~Shahbaz Khan,
  Ming-Hsuan Yang, and Ling Shao.
\newblock Cycleisp: Real image restoration via improved data synthesis.
\newblock In {\em Proc. IEEE Conf. Comput. Vis. Pattern Recognit.}, pages
  2696--2705, 2020.

\bibitem{Zamir2020MIRNet}
Syed~Waqas Zamir, Aditya Arora, Salman Khan, Munawar Hayat, Fahad~Shahbaz Khan,
  Ming-Hsuan Yang, and Ling Shao.
\newblock Learning enriched features for real image restoration and
  enhancement.
\newblock In {\em Proc. Eur. Conf. Comput. Vis.}, 2020.

\bibitem{zhang2017beyond}
Kai Zhang, Wangmeng Zuo, Yunjin Chen, Deyu Meng, and Lei Zhang.
\newblock Beyond a gaussian denoiser: Residual learning of deep cnn for image
  denoising.
\newblock {\em IEEE Trans. Image Process.}, 26(7):3142--3155, 2017.

\bibitem{zhang2018ffdnet}
Kai Zhang, Wangmeng Zuo, and Lei Zhang.
\newblock Ffdnet: Toward a fast and flexible solution for cnn-based image
  denoising.
\newblock {\em IEEE Trans. Image Process.}, 27(9):4608--4622, 2018.

\bibitem{zhang2010two}
Lei Zhang, Weisheng Dong, David Zhang, and Guangming Shi.
\newblock Two-stage image denoising by principal component analysis with local
  pixel grouping.
\newblock {\em Pattern Recognit.}, 43(4):1531--1549, 2010.

\bibitem{zhang2021hyperspectral}
Tao Zhang, Ying Fu, and Cheng Li.
\newblock Hyperspectral image denoising with realistic data.
\newblock In {\em Proc. IEEE Int. Conf. Comput. Vis.}, pages 2248--2257, 2021.

\bibitem{zhang2022joint}
Yong Zhang, Wanjie Sun, and Zhenzhong Chen.
\newblock Joint image demosaicking and denoising with mutual guidance of color
  channels.
\newblock {\em Signal Process.}, 200:108674, 2022.

\bibitem{zhang2014novel}
Zemin Zhang, Gregory Ely, Shuchin Aeron, Ning Hao, and Misha Kilmer.
\newblock Novel methods for multilinear data completion and de-noising based on
  tensor-svd.
\newblock In {\em Proc. IEEE Conf. Comput. Vis. Pattern Recognit.}, pages
  3842--3849, 2014.

\bibitem{zoran2011learning}
Daniel Zoran and Yair Weiss.
\newblock From learning models of natural image patches to whole image
  restoration.
\newblock In {\em Proc. IEEE Int. Conf. Comput. Vis.}, pages 479--486, 2011.

\end{thebibliography}
}

\end{document}